\documentclass [12pt]{article}
\usepackage{amsmath,amssymb,cite}
\usepackage{appendix}
\usepackage{feynmp}
\usepackage{float}
\usepackage[vcentermath,enableskew]{youngtab}
\usepackage{tabularx}
\usepackage[english]{babel}
\usepackage{graphicx}
\usepackage{indentfirst}
\usepackage{epsfig}
\usepackage{epstopdf}
\usepackage[]{hyperref}
\usepackage[section]{placeins}
\usepackage[stable]{footmisc}
\usepackage{appendix}
\usepackage{tabularx}
\usepackage[english]{babel}
\usepackage{graphicx}
\usepackage{indentfirst}
\usepackage{epsfig}
\usepackage{slashed}
\usepackage{fancyhdr}
\numberwithin{equation}{section}

\fancypagestyle{plain}{\fancyhead{}
}

\begin{document}

\title{
  Two approaches to quantum gravity and M-(atrix) theory at large number of dimensions
}

\author{Badis Ydri\\
Department of Physics, Annaba University,\\
 Annaba, Algeria.
}

\maketitle

\begin{abstract}
  A Gaussian approximation to the bosonic part of M-(atrix) theory with mass deformation is considered at large values of the dimension $d$. From the perspective of the gauge/gravity duality this action reproduces with great accuracy  the stringy Hagedorn phase transition from a confinement (black string) phase to a deconfinement (black hole) phase whereas from the perspective of the matrix/geometry approach this action only captures a remnant of the  geometric Yang-Mills-to-fuzzy-sphere phase where the fuzzy sphere solution is only manifested as a three-cut configuration termed the  "baby fuzzy sphere" configuration. The Yang-Mills phase retains most of its characteristics with two exceptions: i) the uniform distribution inside a solid ball suffers a crossover at very small values of the gauge coupling constant to a Wigner's semi-circle law, and ii) the uniform distribution at small values of the temperatures is non-existent.

\end{abstract}

\tableofcontents

\section{Introduction and results}
M-(atrix) theory can be thought of as a "grand unified theory" of the following two main approaches to quantum gravity:
\begin{itemize}
\item {\bf The gauge-theory/Einstein-gravity duality:}  This is the dominant and most successful current approach to quantum gravity with the celebrated holographic correspondence (also known as the AdS/CFT correspondence) as the primary example \cite{Maldacena:1997reF,Gubser:1998bcF,Witten:1998qjF}.
  \begin{itemize}
  \item Indeed, M-(atrix) theory appeared first (as far as I know) as the action for the quantized membrane \cite{deWit:1988wriF}. This action admits supergravity in $11$ dimensions as a low energy limit and thus  can provide the UV completion of the $11-$dimensional M-theory (hence the name M-(atrix) theory) which unifies all five superstring theories in $10$ dimensions and provides their non-perturbative formulation.
    
  \item The M-(atrix) theory was then suggested as the action for the low energy dynamics  of D0-branes (in type IIA string theory). This is the famous BFSS matrix quantum mechanics model \cite{Banks:1996vhF}. The large $N$ limit of this model (where $N$ is the number of D0-branes) is conjectured to describe M-theory in the infinite momentum frame in the uncompactified limit (or equivalently its light-cone quantization). The Hamiltonian is given in terms of $N\times N$ hermitian matrices $X^i$ and $P^i$ by
    \begin{eqnarray}
      H=Tr\bigg[\frac{1}{2}P^iP^i-\frac{g^2}{4}[X^i,X^j]^2+g\Psi\gamma^i[X^i,\Psi]\bigg].\label{Ham}
    \end{eqnarray}
    The Yang-Mills term provides therefore the potential for the generalized coordinates $X^i$ whereas the generalized momenta $P^i$ give the kinetic term. The hermitian spinorial matrices $\Psi_{\alpha}$ yield the fermionic correction to the Yang-Mills potential.
  \item This M-(atrix) theory is another example of the gauge-theory/Einstein-gravity duality which is quite distinct from the much more studied  AdS/CFT correspondence. The gravity dual in this case is the black 0-brane which is a $10-$dimensional black hole solution in type IIA supergravity \cite{Horowitz:1991cdF,Gibbons:1987psF,Itzhaki:1998ddF}.
  \item Although M-(atrix) theory is not strictly speaking an AdS/CFT correspondence it is intimately connected to the case AdS$_3$/CFT$_2$ \cite{Aharony:2004igF}. For example, in $1+1$ dimensions the black-string/black-hole phase transition (which is an example of the Gregory-Laflamme instability \cite{Gregory:1993vyF}) can be successfully studied at high temperature using M-(atrix) theory and its black 0-brane configuration whereas at low temperature we can employ the holographic method using the AdS$_3$/CFT$_2$.
    \item Again, although M-(atrix) theory is not strictly speaking an AdS/CFT correspondence the interpretation of spacetime geometry in terms of entanglement entropy is also possible using a generalized connection between areas and entropies \cite{Anous:2019rqbF} similar to the Ryu-Takayanagi formula \cite{Ryu:2006bvF}.

    \end{itemize}
\item {\bf The matrix-model/noncommutative-geometry approach:} This is the noncommutative geometry approach to quantum gravity infused with the matrix model approaches to string theory.
  \begin{itemize}
     \item
      The dimensional reduction of $(d+1)-$dimensional ${\cal N}=1$ supersymmetric Yang-Mills theory to $(p+1)-$dimension describes the dynamics of Dp-branes in the low energy
      limit \cite{Witten:1995imF,Polchinski:1995mtF}. These low-dimensional gauge theories contain $d-p$ adjoint scalar fields $\Phi_i$ playing the role of the spacetime coordinate operators, i.e. $\Phi_i=X^i$.

      In particular, the dimensional reduction to $1-$dimension corresponding to D0-branes is precisely the M-(atrix) theory or the BFSS matrix quantum mechanics. This $1-$dimensional action reads 

\begin{eqnarray}
S=\frac{1}{g^2}\int_0^{\beta}dt{\rm Tr}\bigg[\frac{1}{2}(D_t\Phi_i)^2-\frac{1}{4}[\Phi_i,\Phi_j]^2+\frac{1}{2}\Psi_{\alpha}D_t\Psi_{\alpha}-\frac{1}{2}\Psi_{\alpha}(\gamma_i)_{\alpha\beta}[\Phi_i,\Psi_{\beta}]\bigg].\label{BFSS0}
\end{eqnarray}
The only free parameter in this model is the Hawking temperature $T=1/\beta$. 
\item The dimensional reduction to $0-$dimension gives the celebrated IKKT matrix model where the bosonic term in the action only involves the Yang-Mills term \cite{Ishibashi:1996xsF}.  Indeed, the compactification of M-(atrix) theory on a circle gives us the IKKT matrix model which is perhaps the most fundamental of all matrix models. This involves essentially setting the kinetic term in the Hamiltonian (\ref{Ham}) to zero which corresponds in turn to the action
  \begin{eqnarray}
S=-\frac{1}{4g^2}[\Phi_i,\Phi_j]^2+Tr\Psi\gamma^i[\Phi_i,\Psi].
\end{eqnarray}
  \item By combining the principles of quantum mechanics and general relativity we are led to the picture that spacetime geometry below the Planck scale has no operational meaning \cite{Doplicher:1994tuF}. This leads immediately to the quantum commutation relations ($\lambda_P$ is the Planck length)
    \begin{eqnarray}
      [x_{\mu},x_{\nu}]=i\lambda_P^2 Q_{\mu\nu}.
    \end{eqnarray}
    This solution is the classical background of the action principle 
    \begin{eqnarray}
      S=\frac{1}{4g^2}Tr\bigg(i[x_{\mu},x_{\nu}]+\lambda_P^2 Q_{\mu\nu}\bigg)^2.
    \end{eqnarray}
    This action contains the Yang-Mills term as a central piece and should be compared with the IKKT matrix model. In fact, this action is a mass deformation of (a sort of) the IKKT matrix model. However, the variables here are the coordinates $x_{\mu}$ and the noncommutativity tensor $Q_{\mu\nu}$ which satisfies some constraints resulting therefore in extra dimensions to the Minkwoski spacetime  $M^4$ given essentially by ${\bf S}^2\otimes {\bf S}^2$.
  \item The much studied Moyal-Weyl spacetime corresponds to a constant noncommutativity tensor, i.e. $Q_{\mu\nu}=\theta_{\mu\nu}$ which clearly breaks Lorentz symmetry but removes the need for the extra dimensions ${\bf S}^2\otimes {\bf S}^2$. As it turns out, the dynamics  of open strings moving in a flat space in the presence of a non-vanishing Neveu-Schwarz B-field   and with Dp-branes is equivalent to leading order in the string tension to a gauge theory on  a Moyal-Weyl space ${\bf R}^d_{\theta}$ \cite{Seiberg:1999vsF}.
  \item Emergent (noncommutative) geometry can be obtained from M-(atrix) theory (or equivalently from its compactification on the circle given by the IKKT matrix model) by adding to the action mass deformation terms such as the Myers term  \cite{Myers:1999psF}. The resulting actions are closely related to the BMN matrix model which differs from the BFSS matrix model by the plane wave deformation and thus preserves maximal supersymmetry \cite{Berenstein:2002jqF}.

 \item Noncommutative quantum gravity based on Poisson manifolds can also be embedded successfully in the IKKT matrix model and M-(atrix) theory. This emergent gravity (which also features black hole solutions) should then be thought of as the dual gravity theory to these matrix models within this approach.  See \cite{Steinacker:2010rhF} for an elementary introduction.
    \end{itemize}
\end{itemize}

In this note our scope is modest and we will only concern ourselves with the bosonic part of the M-(atrix) theory action (\ref{BFSS0}) at large values of the dimension $d$ with ($\alpha\ne 0$) and without ($\alpha=0$) mass deformation given by the Chern-Simons term (Myers effect). The bosonic action for finite $d$ reads
\begin{eqnarray}
S_{\alpha,d}=\frac{1}{g^2}\int_0^{\beta}dt Tr\bigg[\frac{1}{2}(D_t\Phi_a)^2-\frac{1}{4}[\Phi_a,\Phi_b]^2+\frac{2i\alpha}{3}\epsilon_{abc}\Phi_a\Phi_b\Phi_c\bigg].\label{ter}
\end{eqnarray}
The physics of this theory  is already extremely rich  \cite{Kawahara:2007fnF,Kawahara:2007nwF}.

From the perspective of the gauge/gravity duality this action exhibits in the limit $\alpha\longrightarrow 0$ a Hagedorn (first or second order) phase transition from a uniform (confinement, black string) phase to a non-uniform (deconfinement, black hole) phase closely followed by a Gross-Witten-Wadia (third order) transition from a gapless phase to a gapped (black hole) phase.

Whereas from the perspective of the matrix/geometry approach the above action exhibits in the limit $\alpha\longrightarrow \infty$ a geometric (first and second order characters) transition from a Yang-Mills phase to a fuzzy sphere \cite{HoppeF,Madore:1991bwF} phase which exists always in the deconfinement (black hole) phase. This bosonic BFSS geometric phase enjoys the same characteristics as the bosonic IKKT Yang-Mills-to-fuzzy-sphere phase. See \cite{DelgadilloBlando:2007vxF,Azuma:2004zqF,CastroVillarreal:2004vhF} and references therein.

In the remainder of the phase diagram between small $\alpha$ and large $\alpha$ there is a smooth interpolation between the two sets of physics.

The goal in this article is to study this phase diagram in the large $d$ limit of the model (\ref{ter}). It is known that the bosonic part of the M-(atrix) theory action (\ref{BFSS0}) without mass deformation is approximated in the large $d$ limit by the Gaussian model \cite{Mandal:2009vzF,Mandal:2011hbF} 

\begin{eqnarray}
S_{\alpha=0,d=\infty}=\frac{1}{g^2}\int_0^{\beta} dt Tr\bigg[\frac{1}{2}(\partial_t\Phi_i)^2+\frac{1}{2}m^2(\Phi_i)^2\bigg].\label{ter1}
\end{eqnarray}
The mass $m$ is related to the dimension by $m=d^{1/3}$. Thus, instead of studying the phase diagram of the full bosonic action $S_{\alpha,d}$, which has been done for example in   \cite{Kawahara:2007fnF,Kawahara:2007nwF, Filev:2015hiaF}, we will study the phase diagram of the following cubic action   (considered here in its own right)

\begin{eqnarray}
S_{\alpha,d=\infty}=\frac{1}{g^2}\int_0^{\beta} dt Tr\bigg[\frac{1}{2}(\partial_t\Phi_i)^2+\frac{1}{2}m^2(\Phi_i)^2+\frac{2i\alpha}{3}\epsilon_{abc}\Phi_a\Phi_b\Phi_c\bigg]\label{ter2}.
\end{eqnarray}
This cubic model requires a regularization and we will choose as our regulator term a double-trace potential which has no effect on the physics except the removal of a zero mode from the scalar sector.

As we will see all the characteristics of the  black-hole-to-black-string phase transition are captured to a very good accuracy using the Gaussian model (\ref{ter1}) and the effectively Gaussian model (\ref{ter2}) (since the Chern-Simons term in  (\ref{ter2}) is effectively vanishing in the Yang-Mills phase) whereas only a remnant  of the Yang-Mills-to-fuzzy-sphere phase  is reproduced by means of the cubic action  (\ref{ter2}).

This article is organized as follows. In section $2$ we will study the model (\ref{ter1}) (with $d=9$) and derive by means of the Monte Carlo method all the properties of the   black-hole-to-black-string phase transition.

In section $3$ we will study the model (\ref{ter2})  (with $d=3$ which is still quite large) by means of the Monte Carlo, perturbative and matrix methods  and show that only a remnant of the fuzzy sphere persists which consist of a three-cut solution corresponding to the irreducible representation of $SU(2)$ given by the direct sum $0 \oplus \frac{1}{2}$. This configuration is termed in this article a "baby fuzzy sphere" solution. As it turns out, this solution is always existing  in the deconfining phase of the gauge theory.

Furthermore, it is observed that the Yang-Mills phase becomes divided into two distinct regions with a crossover between them. The first  region is dominated by the Wigner's semi-circle law at very low values of the gauge coupling constant $\tilde{\alpha}$ and a second region inside the Yang-Mills phase dominated by a uniform distribution occurring at medium values of the gauge coupling constant before reaching the baby fuzzy sphere boundary. Here "uniform distribution" refers to the coordinate matrices $\Phi_i$ and not to the holonomy matrix and Polyakov line. At low temperatures the middle phase (uniform distribution) is very narrow and could be non-existent. The boundary between the two phases (Yang-Mills and baby fuzzy sphere phases) is constructed numerically and analytically and it is shown that the scaling of the gauge coupling constant and the temperature  in the two phases is possibly different resulting in the fact that either the geometric baby fuzzy sphere phase removes the uniform confining phase (which is the most plausible physical possibility) or the other way around, i.e. the gauge theory uniform confining phase removes the baby fuzzy sphere phase.   

We conclude in section $4$ with a summary of the main/current (Monte Carlo) results and an outlook.

\section{The Hagedorn or deconfinement transition at large $d$}
\subsection{The phase structure of the (lattice) bosonic action}
We start then with the bosonic part of the M-(atrix) theory (\ref{BFSS0}) at large values of the dimension $d$. The action reads explicitly
\begin{eqnarray}
S=\frac{1}{g^2}\int_0^{\beta}dt Tr\bigg[\frac{1}{2}(D_t\Phi_i)^2-\frac{1}{4}[\Phi_i,\Phi_j]^2\bigg].\label{BFSST}
\end{eqnarray}
The only free parameter in this model is the Hawking temperature $T=1/\beta$. On the lattice with the time direction given by a circle the inverse temperature $\beta$ is precisely equal to the circumference of the circle, viz $\beta=a.\Lambda$ where $a$ is the lattice spacing and $\Lambda$ is the number of links.

The energy of the bosonic truncation (\ref{BFSST}) of the BFSS matrix model (\ref{BFSS0}) is defined by (where $Z(\beta)$ is the partition function at temperature $T=1/\beta$)

\begin{eqnarray}
E=-\frac{1}{Z(\beta)}\frac{Z(\beta^{'})-Z(\beta)}{\Delta\beta}~,~\Delta\beta=\beta^{'}-\beta\longrightarrow 0.
\end{eqnarray}
We compute immediately \cite{Kawahara:2007fnF}
\begin{eqnarray}
\frac{E}{N^2}=\frac{3T}{N^2}\langle {\rm commu}\rangle~,~{\rm commu}=-\frac{1}{4g^2}\int_{0}^{\beta}dt{Tr}[{\Phi}_i^{},{\Phi}_j^{}]^2.
\end{eqnarray}
The corresponding radius or extent of space $R^2$ is another very important observable in this model which is given explicitly by

\begin{eqnarray}
R^2=\frac{a}{\Lambda N^2}\langle {\rm radius}\rangle~,~{\rm radius}=\frac{N}{a}\sum_{n=1}^{\Lambda}{Tr}{\Phi}_i^{2}(n).
\end{eqnarray}
The Polyakov line (which acts as our macroscopic order parameter) is defined in terms of the holonomy matrix $U$ (or Wilson loop) by  the relation

 \begin{eqnarray}
P=\frac{1}{N}Tr U~,~U={\cal P}\exp(-i\int_0^{\beta}dt A(t)).
\end{eqnarray}
After gauge-fixing on the lattice (we choose the static gauge $A(t)=-(\theta_1,\theta_2,...,\theta_N)/\beta$) we write the Polyakov line $P$ in terms of the holonomy angles $\theta_a$ as

 \begin{eqnarray}
P=\frac{1}{N}\sum_a\exp(i\theta_a).
\end{eqnarray}
We actually measure in Monte Carlo simulation the expectation value

 \begin{eqnarray}
\langle |P|\rangle=\int d\theta \rho(\theta)\exp(i\theta).
\end{eqnarray}
The eigenvalue distribution $\rho(\theta)$ of the holonomy angles is our microscopic order parameter used to characterize precisely the various phases of this model.   This eigenvalue distribution is given formally by

\begin{eqnarray}
\rho(\theta)=\frac{1}{N}\sum_{a=1}^N \langle\delta(\theta-\theta_a)\rangle.
\end{eqnarray}
At high temperatures the bosonic part of the BFSS quantum mechanics reduces to the bosonic part of the IKKT model \cite{Kawahara:2007ibF}. The leading behavior of the various observables of interest at high temperatures can be obtained in terms of the corresponding expectation values in the IKKT model. We get then

\begin{eqnarray}
  R^2=\sqrt{T}\chi_1.\label{cal1}
\end{eqnarray}

\begin{eqnarray}
\langle |P|\rangle=1-\frac{1}{2d}{T}^{-3/2}\chi_1.\label{cal2}
\end{eqnarray}

\begin{eqnarray}
\frac{E}{N^2}=\frac{3}{4}{T}\chi_2~,~\chi_2=(d-1)(1-\frac{1}{N^2}).\label{cal3}
\end{eqnarray}
The coefficient $\chi_1$ for various $d$ and $N$ can be read off from table $1$ of \cite{Kawahara:2007ibF} whereas the coefficient $\chi_2$ was determined exactly from the Schwinger-Dyson equation.

The behavior  given by equations (\ref{cal1}), (\ref{cal2}) and (\ref{cal3}) can be used to calibrate  Monte Carlo simulations at high  temperatures.

The phase diagram of this model was determined numerically by means of the Monte Carlo method in \cite{Kawahara:2007fnF} to be consisting of two phase transitions and three stable phases.

\begin{itemize}
\item {\bf The confinement/deconfinement phase transition:}   At low temperatures the $U(1)$ symmetry $A(t)\longrightarrow A(t)+C.{\bf 1}$ is unbroken and hence we have a confining phase  characterized by a uniform eigenvalue distribution. The $U(1)$ symmetry gets spontaneously broken at some high temperature $T_{c 2}$ and the system  enters the deconfining phase which is characterized by a non-uniform eigenvalue distribution. This is a second order phase transition.

\item {\bf The Gross-Witten-Wadia phase transition:} This is a third order phase transition occurring at a temperature $T_{c1}> T_{c2}$ dividing therefore the non-uniform phase into two distinct phases: The gapless phase in the intermediate region $T_{c 2}\le T\le T_{c 1}$ and the gapped phase at high temperatures $T> T_{c1}$\footnote{The terminology "gapless" means that the eigenvalue distribution has no gaps on the circle whereas "gapped" means that the distribution is non-zero only in the range $[-\theta_0,\theta_0]$}. This transition is well described by the Gross-Witten-Wadia one-plaquette unitary model \cite{Gross:1980heF,Wadia:1980cpF}.

\item {\bf The $1/d$ expansion:}  By using a $1/d$ expansion around the $d=\infty$ saddle point of the model (\ref{BFSST}) which  is characterized by a non-zero value of the condensate $\langle Tr\Phi_i\Phi_i\rangle$ we can show explicitly that the phase structure of the model consists of two phases: $1)$ a confinement/deconfinement second order phase transition marking the onset of non-uniformity in the eigenvalue distribution closely followed by $2)$ a GWW third order phase transition marking the onset of a gap in the eigenvalue distribution \cite{Mandal:2009vzF,Mandal:2011hbF}. See also  \cite{Drouffe:1979dhF,Drouffe:1983fvF,Kabat:2000zvF,Kabat:2001veF,Hotta:1998enF}.

\item {\bf Hagedorn transition:}

  It has been argued that the deconfinement phase transition in gauge theory such as the above discussed phase transition is precisely the Hagedorn phase in string theory \cite{Aharony:2003sxF, AlvarezGaume:2005fvF}. It has also been argued there that the Hagedorn transition could be a single first order transition and not a deconfinement second order transition followed by a gapped third order transition, i.e. the gapless phase may not be there (recall that its range is very narrow).

\item {\bf The black-string/black-hole transition:} Indeed, the confinement/deconfinement phase transition observed in this model, which is the analogue of the confinement/deconfinement phase transition of ${\cal N}=4$ supersymmetric Yang-Mills theory on ${\bf S}^3$, is the weak coupling limit of the black-string/black-hole phase transition observed in the dual garvity theory of two-dimensional Yang-Mills theory.

Thus,  the phase structure of gauge theory in one dimension can also be obtained from considerations of holography and the dual gravitational theory (beside and supplementing the Monte Carlo method and the analytical $1/d$ and $1/N$ expansions). In particular, the thermodynamics of a given phase of Yang-Mills gauge theory can be deduced from the Bekenstein-Hawking thermodynamics of the corresponding charged black  (string or hole) solution \cite{Aharony:2004igF}.

The dimensionless parameters of the  two-dimensional super Yang-Mills gauge theory on the torus ${\bf T}^2$ are  $\tilde{T}\tilde{L}$ and $\tilde{\lambda}\tilde{L}^2$ where $\tilde{\beta}=1/\tilde{T}$ and $\tilde{L}$ are the circumferences of the two cycles of ${\bf T}^2$.

At strong 't Hooft coupling $\tilde{\lambda}\longrightarrow\infty$ and small temperature $\tilde{T}$ it was shown in  \cite{Aharony:2004igF} that the  above $2-$dimensional Yang-Mills theory exhibits a first order phase transition at the value

\begin{eqnarray}
\tilde{T}\tilde{L}=\frac{2.29}{\sqrt{\tilde{\lambda}\tilde{L}^2}}.
\end{eqnarray}
This corresponds in the dual gravity theory side to a transition between the black hole phase (gapped phase) and the black string phase (the uniform  and gapless phases) \cite{Susskind:1997drF}. This black-hole/black-string first order phase transition is the Gregory-Laflamme instability in this case \cite{Gregory:1993vyF}.

\end{itemize}
\subsection{Gaussian approximation at large $d$}

For quenched fermions and large number of dimensions $d\longrightarrow\infty$ the BFSS matrix model is equivalent to a gauged matrix harmonic oscillator problem given by the action  \cite{Mandal:2009vzF,Mandal:2011hbF} 

\begin{eqnarray}
S[\Phi]=\frac{1}{g^2}\int_0^{\beta} dt Tr\bigg[\frac{1}{2}(\partial_t\Phi_i)^2+\frac{1}{2}m^2(\Phi_i)^2\bigg]~,~m=d^{1/3}.\label{gauss}
\end{eqnarray}
This gauged matrix harmonic oscillator is  a thermal field theory in one dimension and thus $t$ must be  an imaginary time, viz $\tilde{t}=-it$ is the real time. As before, the fields are periodic with period $\beta=1/T$ where $T$ is the Hawking temperature, and we will study the system in the t'Hooft limit given by

 \begin{eqnarray}
\lambda=g^2N.
\end{eqnarray}
Again, there seems to be  two independent coupling constants $\lambda$ and $T$. However, $g^2$ can always be rescaled away. By dimensional analysis we find that $\Phi_i$ behaves as inverse length and $\lambda$ as inverse length cubed whereas $T$ behaves as inverse length and as a consequence the dimensionless coupling constant must be given by $\tilde{\lambda}={\lambda}/{T^3}$. We will choose $\lambda=1$, i.e. $g^2=1/N$.

It has been argued in \cite{Filev:2015hiaF} that the dynamics of the bosonic BFSS model is fully dominated by the large $d$ behavior encoded in the above quadratic action.

This has been checked in Monte Carlo simulations where a Hagedorn/deconfinement transition is observed consisting of a second order confinement/deconfinement phase transition closely followed by a GWW third order transition which marks the emergence of a gap in the eigenvalue distribution.

First, we must regularize the theory as before using a standard lattice in the Euclidean time direction with a lattice spacing $a$. Recall also that $\beta=\Lambda.a$ is the period on the discrete circle. The lattice gauge-invariant lattice action is straightforwardly given by 

\begin{eqnarray}
S=\frac{1}{g^2}\sum_{n=1}^{\Lambda}{ Tr}\bigg[\frac{1}{a}\Phi_i^2(n)-\frac{1}{a}U_{n,n+1}\Phi_i(n+1)U_{n+1,n}\Phi_i(n)+\frac{1}{2}m^2a\Phi_i(n)^2\bigg].
\end{eqnarray}
The adopted gauge-fixing condition is effectively the so-called static gauge given by \cite{Kawahara:2007fnF,Filev:2015hiaF}

\begin{eqnarray}
U_{n,n+1}=D_{\Lambda}={\rm diag}(\exp(i\frac{\theta_1}{\Lambda}),...,\exp(i\frac{\theta_N}{\Lambda}))~,~\forall n.
\end{eqnarray}
In terms of the gauge field $A$ this condition reads 

\begin{eqnarray}
A(t)=-\frac{1}{\beta}{\rm diag}(\theta_1,...,\theta_N).
\end{eqnarray}
In summary, we are interested in the total lattice action (includin the Faddeev-Popov gauge-fixing term)

\begin{eqnarray}
S_{\rm total}&=&N\sum_{n=1}^{\Lambda}{ Tr}\bigg[\frac{1}{a}{\Phi}_i^{ 2}(n)+\frac{1}{2}am^2{\Phi}_i^{ 2}(n)\bigg]\nonumber\\
&-&\frac{N}{a}\sum_{n=1}^{\Lambda}{Tr}{\Phi}_i^{}(n)D_{\Lambda}{\Phi}_i^{}(n+1)D_{\Lambda}^{\dagger}-\frac{1}{2}\sum_{a\ne b}\ln\sin^2\frac{\theta_a-\theta_b}{2}.
\end{eqnarray}
The Polyakov  line is the order parameter of the Hagedorn transition in string theory and the deconfinment transition in gauge theory which is associated with the spontaneous breakdown of the $U(1)$ symmetry $A(t)\longrightarrow A(t)+C.{\bf 1}$. The Polyakov line is defined in terms of the holonomy matrix $U$ by  the relation

\begin{eqnarray}
P=\frac{1}{N}Tr U~,~U={\cal P}\exp(-i\int_0^{\beta} dt A(t)).
\end{eqnarray}
We compute \cite{Kawahara:2007fnF}

\begin{eqnarray}
U&=&U_{1,2}U_{2,3}...U_{\Lambda-1,\Lambda}U_{\Lambda,1}=D_{\Lambda}^{\Lambda}={\rm diag}(\exp(i\theta_1),...,\exp(i\theta_N).
\end{eqnarray}
Hence

\begin{eqnarray}
P=\frac{1}{N}\sum_ae^{i\theta_a}.
\end{eqnarray}
Another important observable is the radius (or extent of space or more precisely the extent of the eigenvalue distribution) which is  defined by

\begin{eqnarray}
R^2=\frac{a}{\Lambda N^2}\langle {\rm radius}\rangle~,~{\rm radius}=\frac{N}{a}\sum_{n=1}^{\Lambda}{Tr}{\Phi}_i^{2}(n).
\end{eqnarray}
The energy in the present model is given effectively by the extent of space. Indeed, we compute the energy

\begin{eqnarray}
\frac{E}{N^2}=\frac{a^2Tm^2}{N^{2}}\langle{\rm radius}\rangle=m^2R^2.
\end{eqnarray}
We also measure the eigenvalues distribution of the holonomy matrix $U$ and the eigenvalues distribution of the bosonic matrices $\Phi_i(n)$.

\subsection{Preliminary results at large $d$}

We employ the Metropolis algorithm where the variation of the holonomy angles $\theta_a$ requires a very careful treatment due to the center of mass $\theta_{\rm cm}=\sum_{a=1}^N\theta_a/N$ which does not appear in the action and thus behaves as a random walker. See for example \cite{code_Hanada}.

\begin{figure}[htbp]
\begin{center}
  \includegraphics[width=10cm,angle=-0]{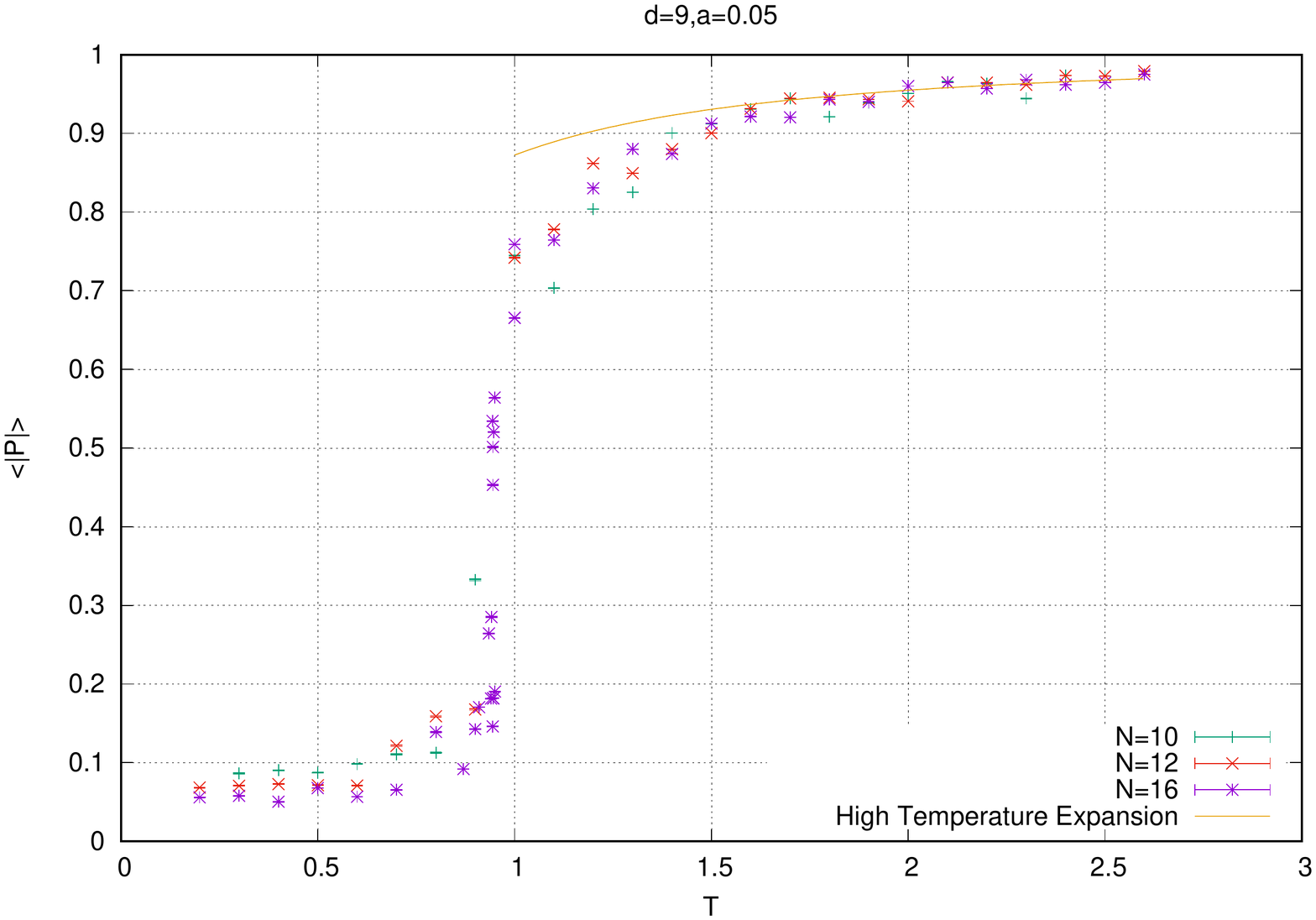}
  \includegraphics[width=10cm,angle=-0]{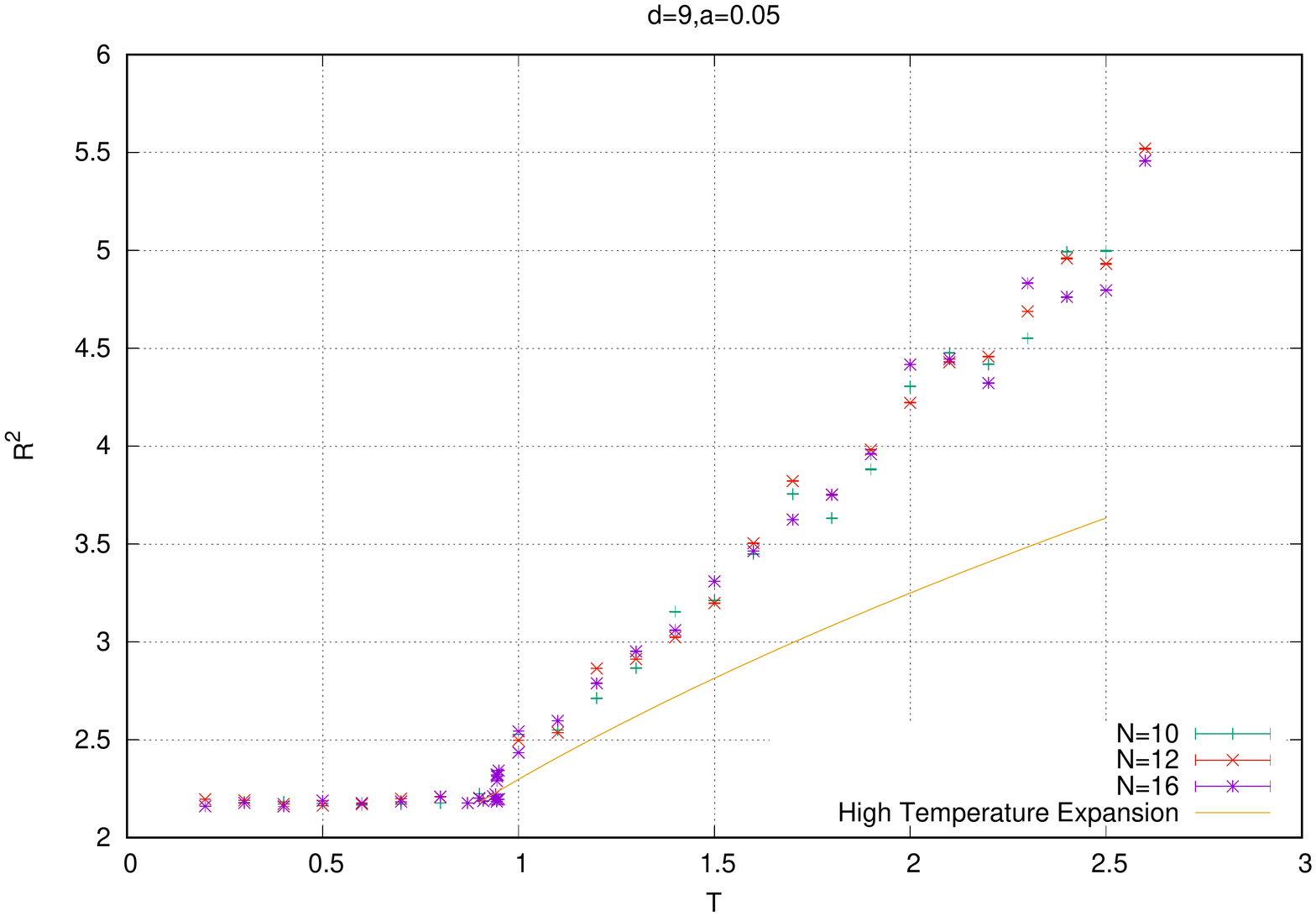}
   \includegraphics[width=10cm,angle=-0]{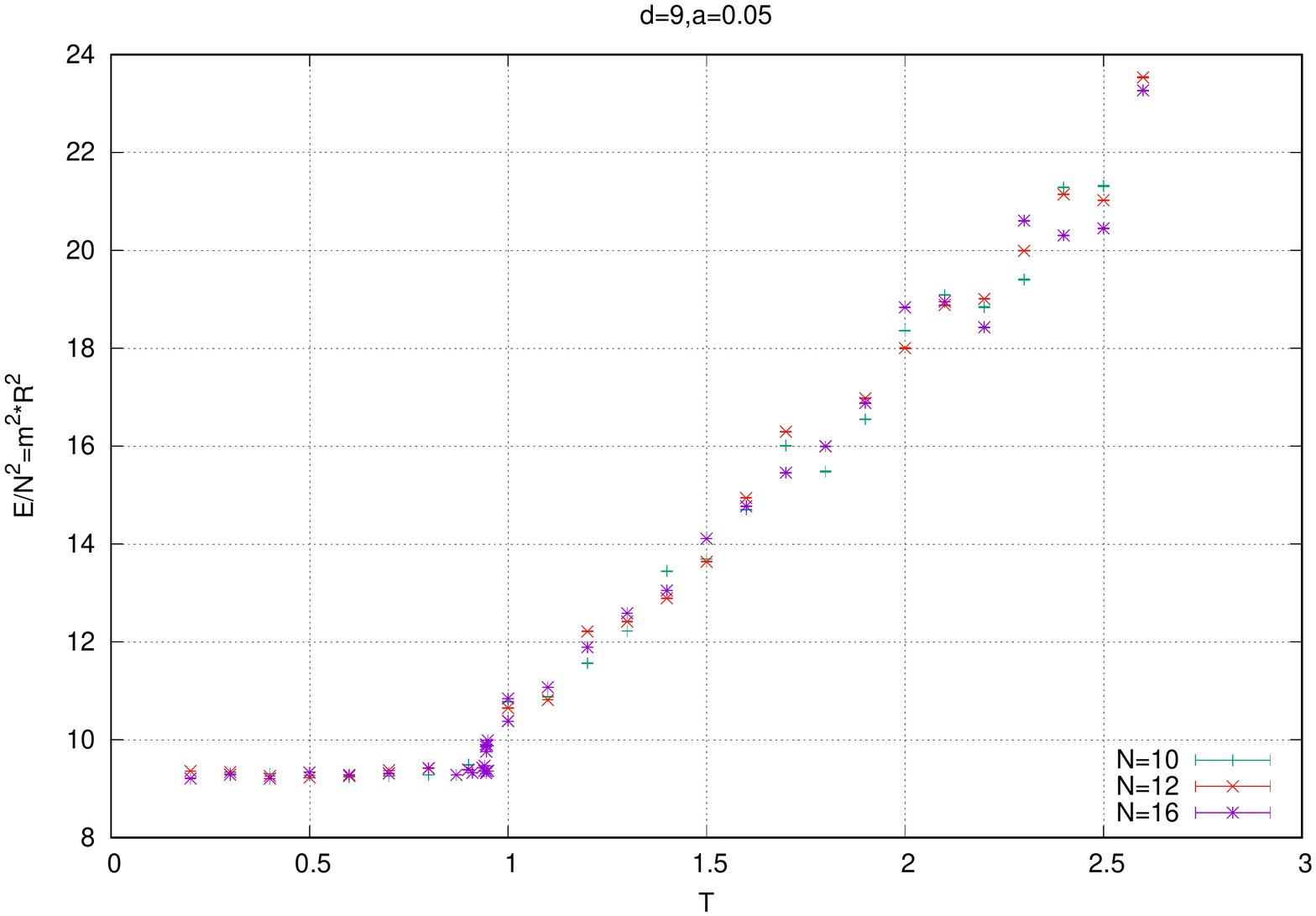}
\end{center}
\caption{The Polyakov line and the extent of space for $N=16$ at large number of dimensions ($d=9$) for lattice $a=0.05$.}\label{sampleold1}
\end{figure}

\begin{figure}[htbp]
\begin{center}
  \includegraphics[width=10cm,angle=-0]{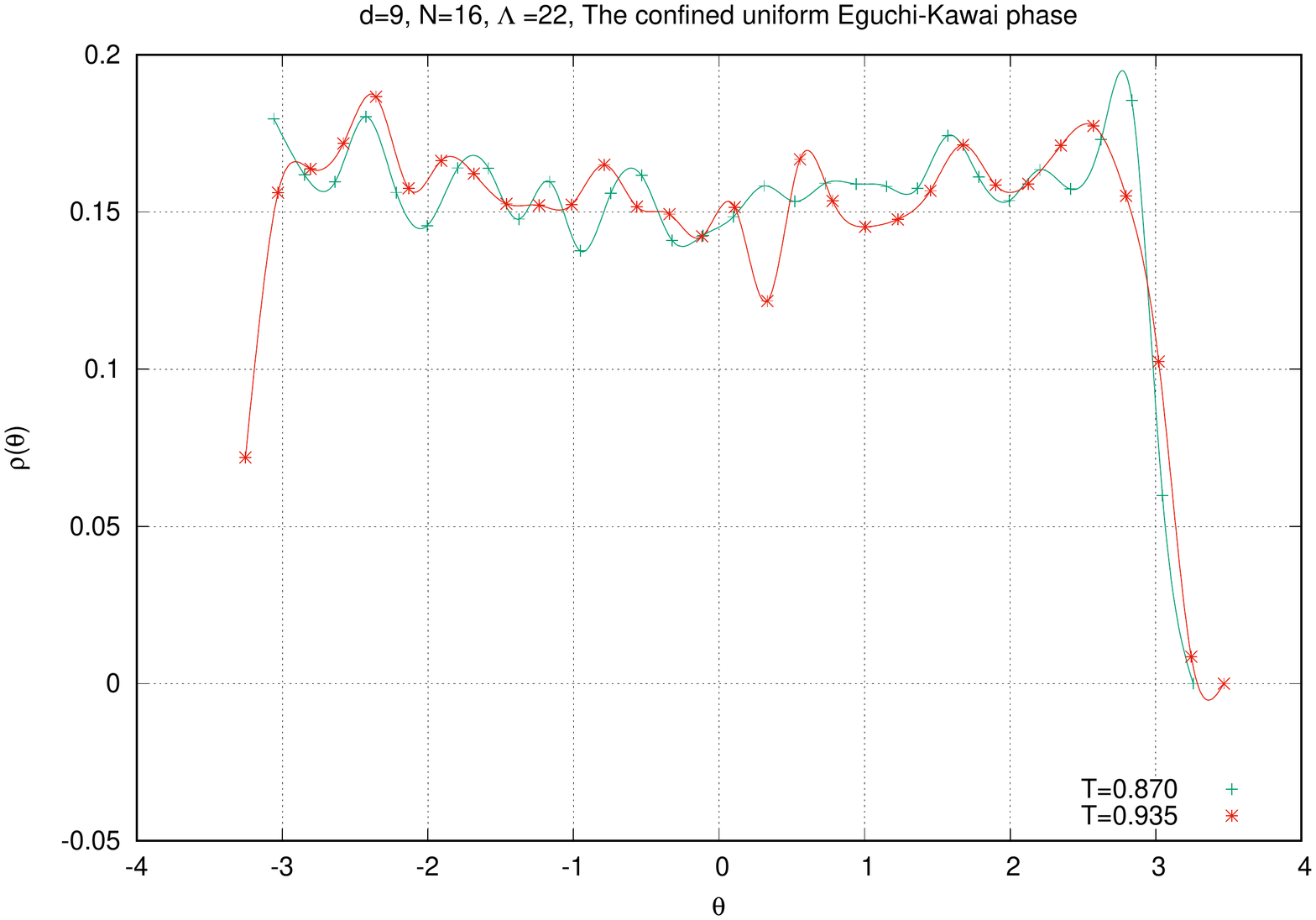}
  \includegraphics[width=10cm,angle=-0]{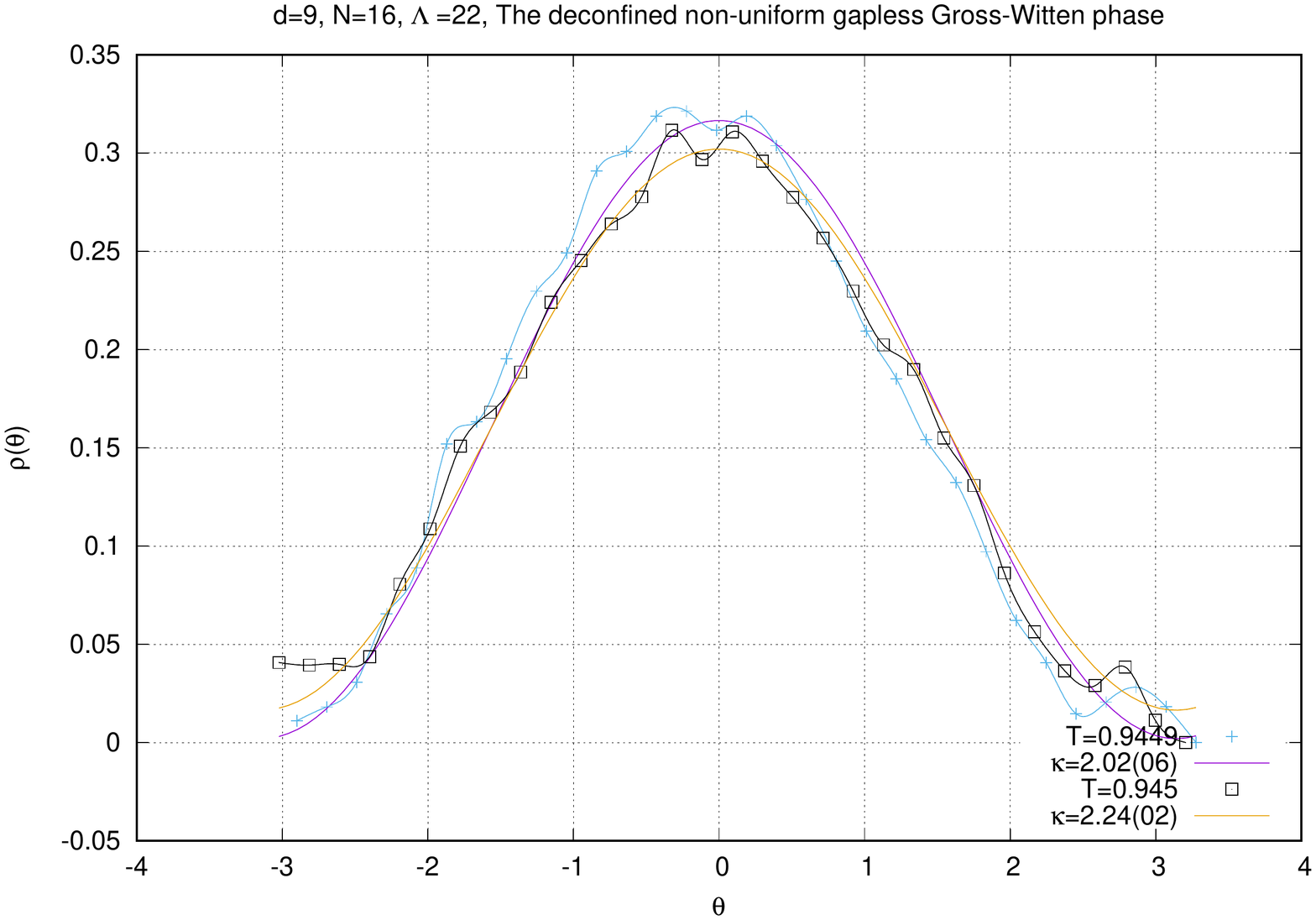}
  \includegraphics[width=10cm,angle=-0]{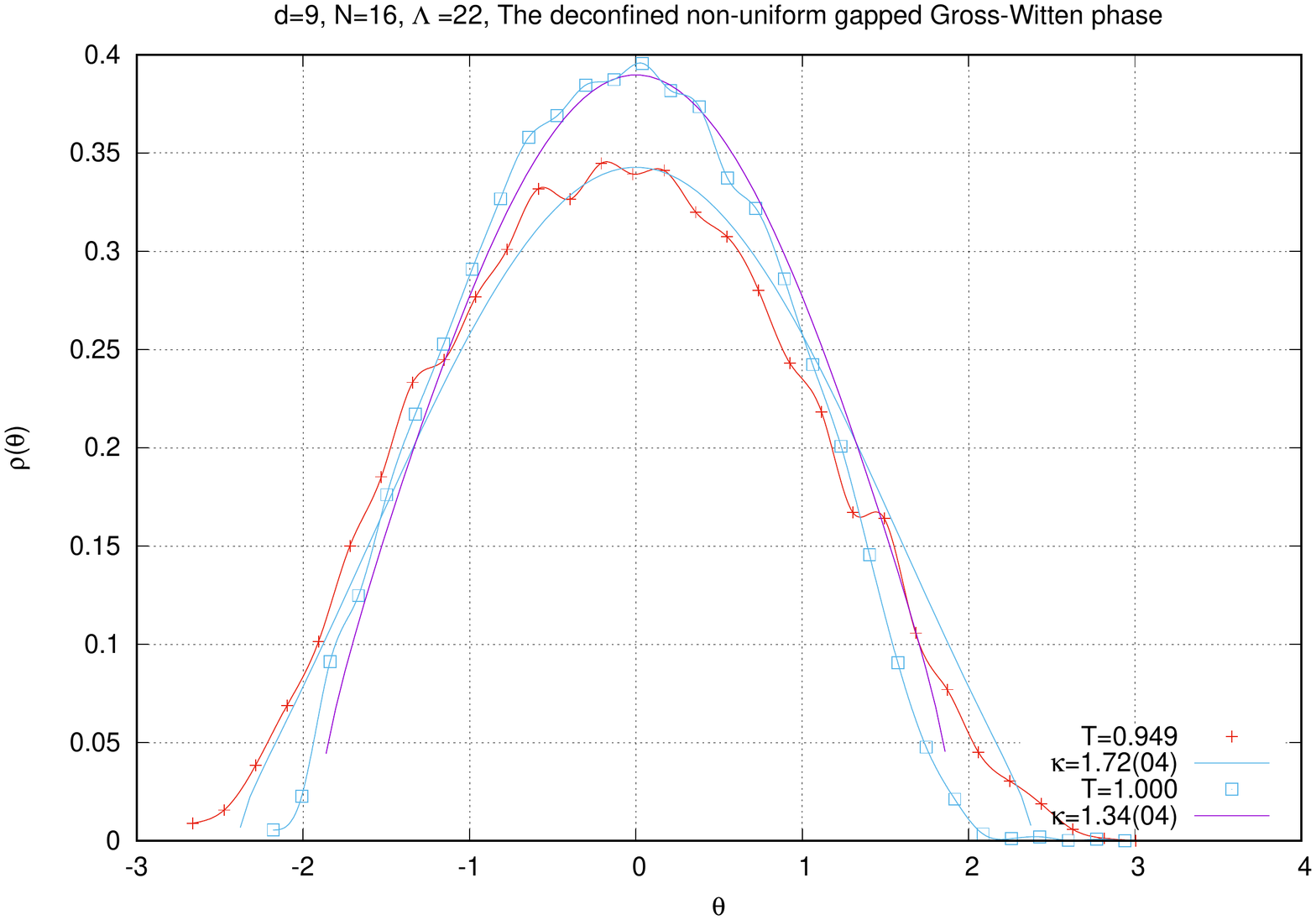}
\end{center}
\caption{The holonomy eigenvalue distribution for $N=16$ at large number of dimensions ($d=9$) for lattice $\Lambda=22$.}\label{sampleold2}
\end{figure}

\begin{figure}[htbp]
\begin{center}
   \includegraphics[width=15cm,angle=-0]{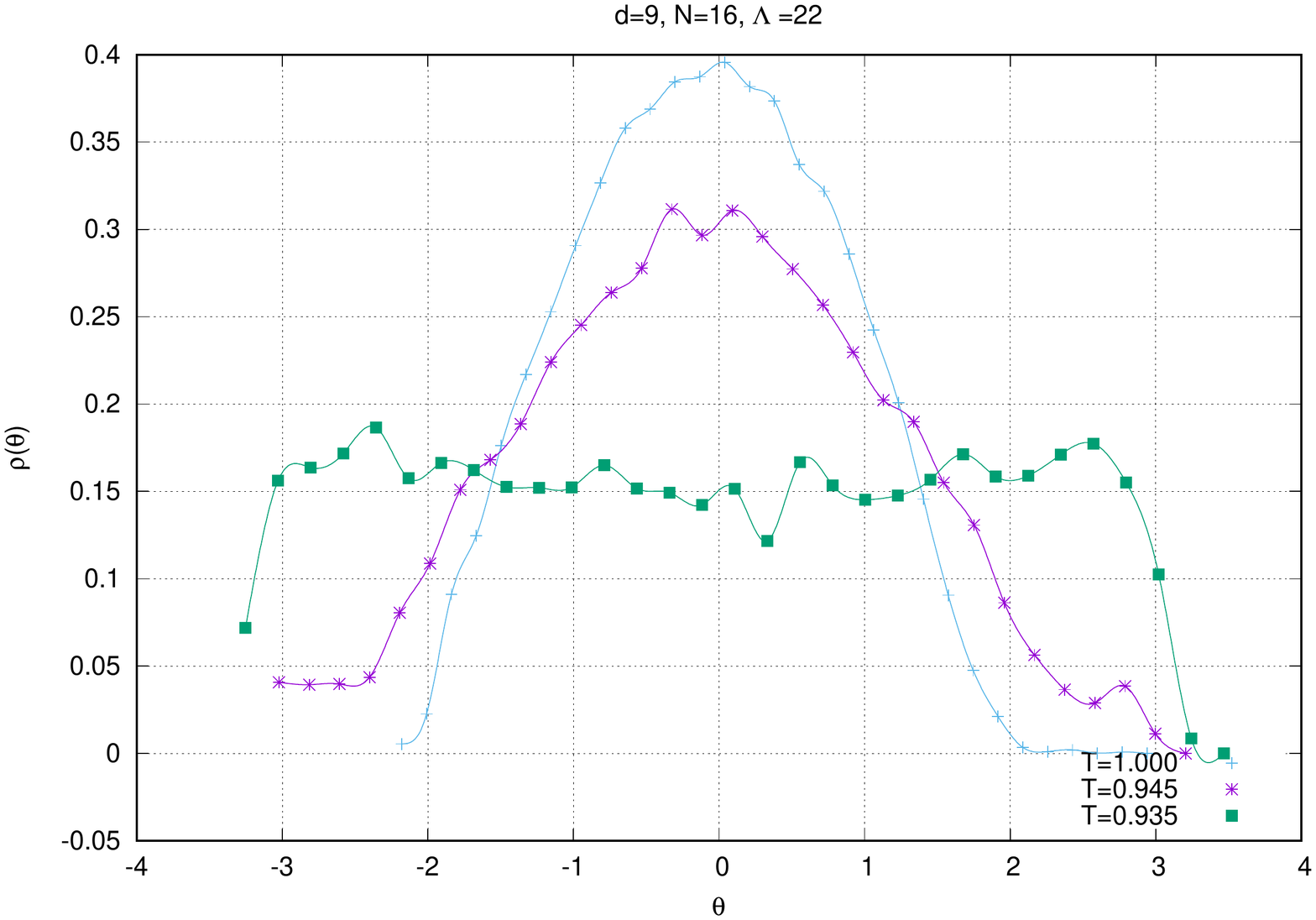}
\end{center}
\caption{The  uniform-to-non-uniform and gapless-to-gapped phase transitions for $N=16$ at large number of dimensions ($d=9$) for lattice $\Lambda=22$.}\label{sampleold3}
\end{figure}

Some of our Monte Carlo results are:

\begin{itemize}

\item First, we can easily check that the Polyakov line $\langle |P|\rangle$ vanishes identically in a uniform eigenvalue distribution.  But numerically it is observed that  $\langle |P|\rangle$ vanishes only as $1/N$ at low temperatures.

See figure (\ref{sampleold1}).

\item The extent of space $R^2$ (or equivalently the energy) in the confining uniform phase is constant which is consistent with the so-called Eguchi-Kawai equivalence \cite{Eguchi:1982nmF}. This states that the expectation values of single-trace operators in $d-$dimensional large $N$ gauge theories are independent of the volume if the $U(1)^d$ symmetry is not spontaneoulsy broken. In our case $d=1$ and independence of the volume is precisely independence of the temperature which is the inverse Euclidean time.

The constant value of the energy in the confining uniform phase is identified with the ground state energy. The energy in the deconfining non-uniform phase ($T> T_{c2}$) deviates from this constant value quadratically, i.e. as  $(T-T_{c 2})^2$.  This is confirmed in the full bosonic model (\ref{BFSST}) in \cite{Kawahara:2007fnF}.

This result is shown on figure (\ref{sampleold1}).

\item It is observed  in numerical simulations that the second phase transition in the non-uniform phase is well described by the Gross-Witten-Wadia one-plaquette model \cite{Gross:1980heF,Wadia:1980cpF}

\begin{eqnarray}
Z_{GWW}=\int dU \exp(\frac{N}{\kappa}Tr U+{\rm h.c}).
\end{eqnarray}
The deconfined non-uniform gapless phase is described by a gapless eigenvalue distribution (and hence the name: gapless phase) of the form

\begin{eqnarray}
\rho_{\rm gapless}=\frac{1}{2\pi}(1+\frac{2}{\kappa}\cos\theta)~,~-\pi<\theta\le+\pi~,~\kappa\ge 2.
\end{eqnarray}
This solution is valid only for $\kappa\ge 2$ where $\kappa$ is a function of the temperature.

At $\kappa=2$ (corresponding to $T=T_{c1}$) a third order phase transition occurs to a gapped eigenvalue distribution given explicitly by

\begin{eqnarray}
\rho_{\rm gapped}=\frac{1}{\pi\sin^2\frac{\theta_0}{2}}\cos\frac{\theta}{2}\sqrt{\sin^2\frac{\theta_0}{2}-\sin^2\frac{\theta}{2}}~,~-\theta_0\le\theta\le+\theta_0~,~\kappa< 2.
\end{eqnarray}
The eigenvalue distribution is non-zero only in the range $[-\theta_0,\theta_0]$ (arbitrarily chosen to be centered around $0$ for simplicity) where the angle $\theta_0$ is given explicitly by

\begin{eqnarray}
\sin^2\frac{\theta_0}{2}=\frac{\kappa}{2}.
\end{eqnarray}
This is a gapped distribution since only the interval $[-\theta_0,\theta_0]$ is filled. At high temperatures corresponding to $\kappa\longrightarrow 0$ the above distribution approaches a delta function \cite{Aharony:2003sxF}.

A sample of these eigenvalue distributions is shown on figures (\ref{sampleold2}) and (\ref{sampleold3}).



\item Also, in this Gaussian approximation it is observed that the eigenvalues of the adjoint scalar fields $\Phi_i$ are distributed according to the Wigner semi-circle law with a radius $r$ following the temperature behavior of  the extent of space $R^2$ since $r^2=4R^2/d$. Thus,  only the radius of the eigenvalue distribution undergoes a phase transition not in its shape (which is always a Wigner semi-circle law). At low temperature this radius becomes constant given by $r=\sqrt{2/m}$.
\end{itemize}

An analytic study of the Gaussian model (\ref{gauss}) is given in the very interesting paper \cite{Furuuchi:2003syF} where its relevance to the plane wave matrix model and string theory is discussed at length.

\section{Emergent geometry in the Gaussian approximation}
\subsection{The Chern-Simons term}
For simplicity, we consider Yang-Mills gauge theory in one dimension with just $d=3$ adjoint scalar fields $\Phi_a$ with an additional Chern-Simons term, viz

\begin{eqnarray}
S=N\int_0^{\beta}dt Tr\bigg[\frac{1}{2}(D_t\Phi_a)^2-\frac{1}{4}[\Phi_a,\Phi_b]^2+\frac{2i\alpha}{3}\epsilon_{abc}\Phi_a\Phi_b\Phi_c\bigg].
\end{eqnarray}
This model was studied at length in \cite{Kawahara:2007nwF}. The basic order parameters are as before the Polyakov line $P$ (or the corresponding eigenvalue distribution $\rho(\theta)$ of the holonomy angles) and the radius $R^2$. But we can also now measure the Chern-Simons term. They are given respectively by

 \begin{eqnarray}
\langle |P|\rangle=\int d\theta \rho(\theta)\exp(i\theta).\label{obs1})
\end{eqnarray}

\begin{eqnarray}
R^2=\frac{1}{\Lambda N}\langle\sum_{n=1}^{\Lambda}{Tr}{\Phi}_i^{2}(n)\rangle.\label{obs2}
\end{eqnarray}

\begin{eqnarray}
  {\rm CS}=\frac{1}{N\Lambda}\langle \frac{2i}{3}\epsilon_{abc}\sum_{n=1}^{\Lambda}Tr\Phi_a(n)\Phi_b(n)\Phi_c(n)\rangle.\label{obs3}
\end{eqnarray}
It is well known that the Chern-Simons term is responsible for the emergence or condensation of the classical geometry of a round sphere (here a round sphere $\times$ Euclidean time) \cite{Myers:1999psF,Alekseev:2000fdF,Azuma:2004zqF}. The global minimum of the theory is a fuzzy sphere given by the spin $j=(N-1)/2$ irreducible representation of $SU(2)$, i.e.
\begin{eqnarray}
  \Phi_a=\alpha L_a~,~[L_a,L_b]=i\epsilon_{abc}L_c~,~c_2=L_aL_a=\frac{N^2-1}{4}.
\end{eqnarray}
Expanding around this minimum by writing $\Phi_a=\alpha(L_a+A_a)$ leads to a gauge theory on the fuzzy sphere with a three-component gauge field $A_a$, i.e. we have effectively a two-dimensional gauge field tangent to the sphere coupled to a normal adjoint scalar field $\phi$ given by
\begin{eqnarray}
  \phi=\frac{\Phi_a^2-c_2\alpha^2}{2\alpha^2\sqrt{c_2}}.\label{using}
\end{eqnarray}
This vacuum state is stable for large values of $\tilde{\alpha}$ defined by $ \tilde{\alpha}=\alpha N^{1/3}$ but it decays (the sphere evaporates) as we decrease $\tilde{\alpha}$ to enter the Yang-Mills phase where the vacuum state becomes given by commuting matrices. 

The large $d$ approximation (assuming that $D=3$ is quite large and $\alpha\longrightarrow 0$) gives immediately the action
\begin{eqnarray}
S_0=N\int_0^{\beta}dt Tr\bigg[\frac{1}{2}(D_t\Phi_a)^2+\frac{1}{2}m^2\Phi_a^2+\frac{2i\alpha}{3}\epsilon_{abc}\Phi_a\Phi_b\Phi_c\bigg]~,~m=d^{1/3}.\label{ac0}
\end{eqnarray}
This action is the analogue of the cubic potential $TrM^2+ g_3TrM^3$ which although is not stable (not bounded from below) admits a well defined large $N$ expansion. In order to access (in Monte Carlo simulations) the non-perturbative physics of this cubic matrix model we can regularize it (embed it) in a stable quartic potential of the form $TrM^2+g_3TrM^3+g_4TrM^4$. Indeed, it is well known that the Liouville theory (the $(2,3)-$minimal conformal matter coupled to two-dimensional quantum gravity) associated with the cubic matrix model   is fully/correctly captured by the renormalization group fixed point  $(g_{3*},g_{4*})=(1/432^{1/4},0)$ of the cubic-quartic matrix model  \cite{Higuchi:1994rvF}.

Similarly here, in order to access in Monte Carlo simulations the non-perturbative physics of the matrix quantum mechanics $S_0$ we should regularize it by adding a quartic potential. By recalling the tracelessness condition of the original model  $Tr\Phi_a=0$ and by demanding rotational invariance on the sphere (any single-trace term is a scalar) we conclude that there are only two possibilities we can contemplate: the  single-trace term $Tr(\Phi_a^2)^2$ and the double-trace term $(Tr\Phi_a^2)^2$. [For example the term $Tr\Phi_aTr\Phi_a\Phi_b^2$ is disallowed because of the tracelessness condition whereas the term $Tr\Phi_a\Phi_b Tr\Phi_a\Phi_b$ is disallowed because the individual traces are not rotational scalars].

This is very reminiscent of the multi-trace expansion of non-commutative scalar field theory \cite{OConnor:2007ibgF} where the kinetic term of scalar fields on non-commutative fuzzy spaces is approximated by an expansion (hopping-parameter-like expansion) of the form (with $\Phi=U\Lambda U^{\dagger}$)

\begin{eqnarray}
  \int dU \exp(Tr[L_a,\Phi]^2)&=&\exp\bigg(a_1Tr\Lambda^2+a_2(Tr\Lambda)^2+b_1 (Tr\Lambda^2)^2+b_2Tr\Lambda^2(Tr\Lambda)^2+b_3(Tr\Lambda)^4\nonumber\\
  &+&b_4Tr\Lambda Tr\Lambda^3+b_5Tr\Lambda^4+...\bigg).
\end{eqnarray}
Indeed, the Yang-Mills term is essentially the kinetic term here and the analogy seems to be more than just an analogy.

The addition of the single-trace term $Tr(\Phi_a^2)^2$ will lead to the decoupling of the normal scalar field \cite{CastroVillarreal:2004vhF} whereas the addition of the  double-trace term $(Tr\Phi_a^2)^2$ will only lead to the decoupling of the zero mode of the normal scalar field. Hence, for simplicity  we are going to consider  the action
\begin{eqnarray}
S_M=M\int_0^{\beta}dt (Tr\Phi_a^2-\varphi c_2 N)^2~,~{c}_2=L_aL_a=\frac{N^2-1}{4}.\label{ac1}
\end{eqnarray}
As it turns out, this is very sufficient to regularize the model $S_0$ without essentially altering it. Indeed, in the large $M$ limit we obtain the condition
\begin{eqnarray}
Tr\Phi_a^2=\varphi c_2 N.\label{constraint0}
\end{eqnarray}
As can be checked quite easily this condition (which replaces the tracelessness condition) kills off only the zero mode of the normal scalar field $\phi$. This can be seen by substituting (\ref{using}) in (\ref{ac1}) with $\varphi=\alpha$ and assuming a fuzzy sphere background to obtain
\begin{eqnarray}
S_M=4\alpha^4N^2c_2M\int_0^{\beta}dt \bigg(\int_{{\bf S}^2} d\Omega\phi\bigg)^2.
\end{eqnarray}
The mass of the zero mode of the normal scalar field is seen to be given by $m^2=8\alpha^4N^2c_2M$ which can become small at small values of the gauge coupling constant $\alpha$. But the  fuzzy sphere background is typically absent in that regime. More importantly, this mass becomes suppressed at high temperatures as $m^2\beta=m^2/T$ which could be problematic (see below). This also suggests that the scaling of the mass $M$ with $N$ at high temperatures (taking also into account the scaling of the parameters $\alpha$ and $T$ given in equation (\ref{419}) ) is given by
\begin{eqnarray}
M=\frac{M_0}{N^4}.\label{ol}
\end{eqnarray}
\subsection{A baby fuzzy sphere}
By varying the regularized action $S_0+S_M$ with respect to $\Phi_a$ we obtain the classical equation of motion
\begin{eqnarray}
D_t^2\Phi_a=m^2\Phi_a+i\alpha\epsilon_{abc}[\Phi_b,\Phi_c]+\frac{4M}{N}\Phi_aTr(\Phi_b^2-\varphi c_2).\label{eom}
\end{eqnarray}
The $SU(2)-$like configurations $\Phi_a=\mu J_a\otimes {\bf 1}_{N/n}$, where $J_a$ is the spin $j=(n-1)/2$ irreducible representation of $SU(2)$, are solution if the scale $\varphi$ is chosen such that

\begin{eqnarray}
\varphi=\frac{\hat{c}_2}{c_2}\mu^2+\frac{1}{2Mc_2}(\frac{m^2}{2}-\alpha\mu)~,~\hat{c}_2=J_aJ_a=\frac{n^2-1}{4}.
\end{eqnarray}
The original model (with the full Yang-Mills term) admits the solutions $\mu=\alpha$ with $n=N$ being the global minimum. We choose then $\varphi$ such that $\mu=\alpha$ and $n=N$, i.e.
\begin{eqnarray}
\varphi=\alpha^2+\frac{1}{2Mc_2}\big(\frac{m^2}{2}-\alpha^2\big).\label{ac2}
\end{eqnarray}
In the large $M$ limit (the limit in which the gauge field on the emergent sphere becomes truly tangential) we see that $\varphi\longrightarrow \alpha^2$.  By substituting this expression in the previous equation we obtain the solution
\begin{eqnarray}
\mu=\frac{\alpha}{4M\hat{c}_2}\bigg[1\pm \sqrt{1+8M\hat{c}_2(2Mc_2-1)}\bigg]\longrightarrow \pm \frac{\alpha N}{2\sqrt{\hat{c_2}}}~,~M\longrightarrow\infty.
\end{eqnarray}
For example, the fundamental configuration $\Phi_a=\mu \sigma_a\otimes {\bf 1}_{N/2}/2$ corresponds to a solution $\mu$ given explicitly by 
\begin{eqnarray}
\mu=\frac{\alpha}{3M}\bigg[1\pm \sqrt{1+6M(2Mc_2-1)}\bigg]\longrightarrow \pm \frac{\alpha N}{\sqrt{3}}~,~M\longrightarrow\infty.
\end{eqnarray}
This was found to be the global minimum of a similar model (Chern-Simons term+ a single-trace quartic potential in $D=0$) in \cite{DelgadilloBlando:2008viF}. In the current matrix quantum mechanics the actual global minimum (in the geometric-like phase) is a combination of $\Phi_a=0$ (commuting matrices) and the fundamental configuration $\Phi_a=\mu J_a$ with $J_a=\sigma_a/2$ which is given explicitly by the following embedding
\begin{eqnarray}
\boxed{  \Phi_a=\bigg(\begin{array}{cc}
      \mu (J_a)_{2\times 2} & 0 \\
      0 & 0
      \end{array}\bigg).}\label{conf}
  \end{eqnarray}
This geometric-like phase is actually a three-cut phase where the three cuts correspond to the two large eigenvalues $\pm \mu/2$ (with multiplicity $1$ each) and to the small eigenvalues around $\mu=0$. This is a fuzzy-sphere-like phase because it corresponds in some sense to a fuzzy sphere with three qubits only. As we lower the coupling a transition will eventually occur to a one-cut phase (the weak coupling regime of the so-called Yang-Mills phase) where all eigenvalues are centered around the eigenvalue $\mu=0$.

\begin{figure}[htbp]
\begin{center}
  \includegraphics[width=15cm,angle=-0]{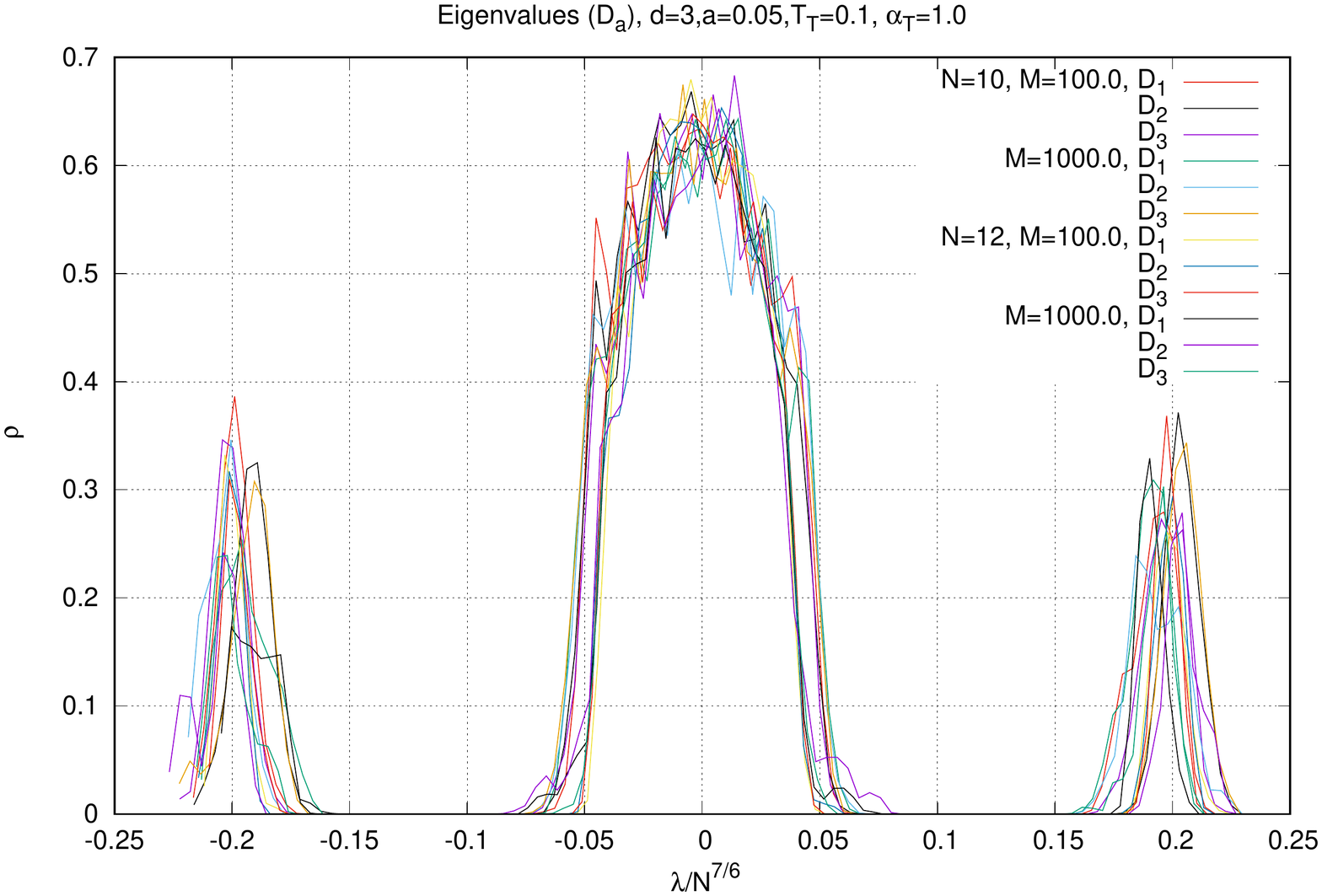}
\end{center}
\caption{The  baby fuzzy sphere configurations for  $\tilde{T}=0.1$ and $\tilde{\alpha}=1.0$ for various values of $N$ and the regularization mass $M$.}\label{sample0v1}
\end{figure}

This three-cut phase is termed in this article a "baby fuzzy sphere phase" because unlike the Yang-Mills phase it is a geometric phase with all the characteristics of the fuzzy sphere phase observed in the original exact model but it comes only with three cuts or qubits, i.e. the direct sum of the irreducible representations $0$ and $1/2$ of $SU(2)$. A sample of this three-cut phase or baby fuzzy sphere phase is shown in figure (\ref{sample0v1}) for temperature $\tilde{T}=0.1$ and value of the gauge coupling constant $\tilde{\alpha}=1$\footnote{In this section and the next we use the value $\tilde{T}=0.1$. We run the Metropolis algorithm for $2^{13} + 2^{13}$ steps where the first $2^{13}$ steps are discarded whereas
  histograms and averages are computed using the last $2^{13}$ steps. We choose one lattice $a=0.05$ corresponding to $\Lambda=35$.}. The limit $M\longrightarrow \infty$ is established quite simply by considering values of $M$ $\geq 100$.

\begin{figure}[htbp]
\begin{center}
  \includegraphics[width=10cm,angle=-0]{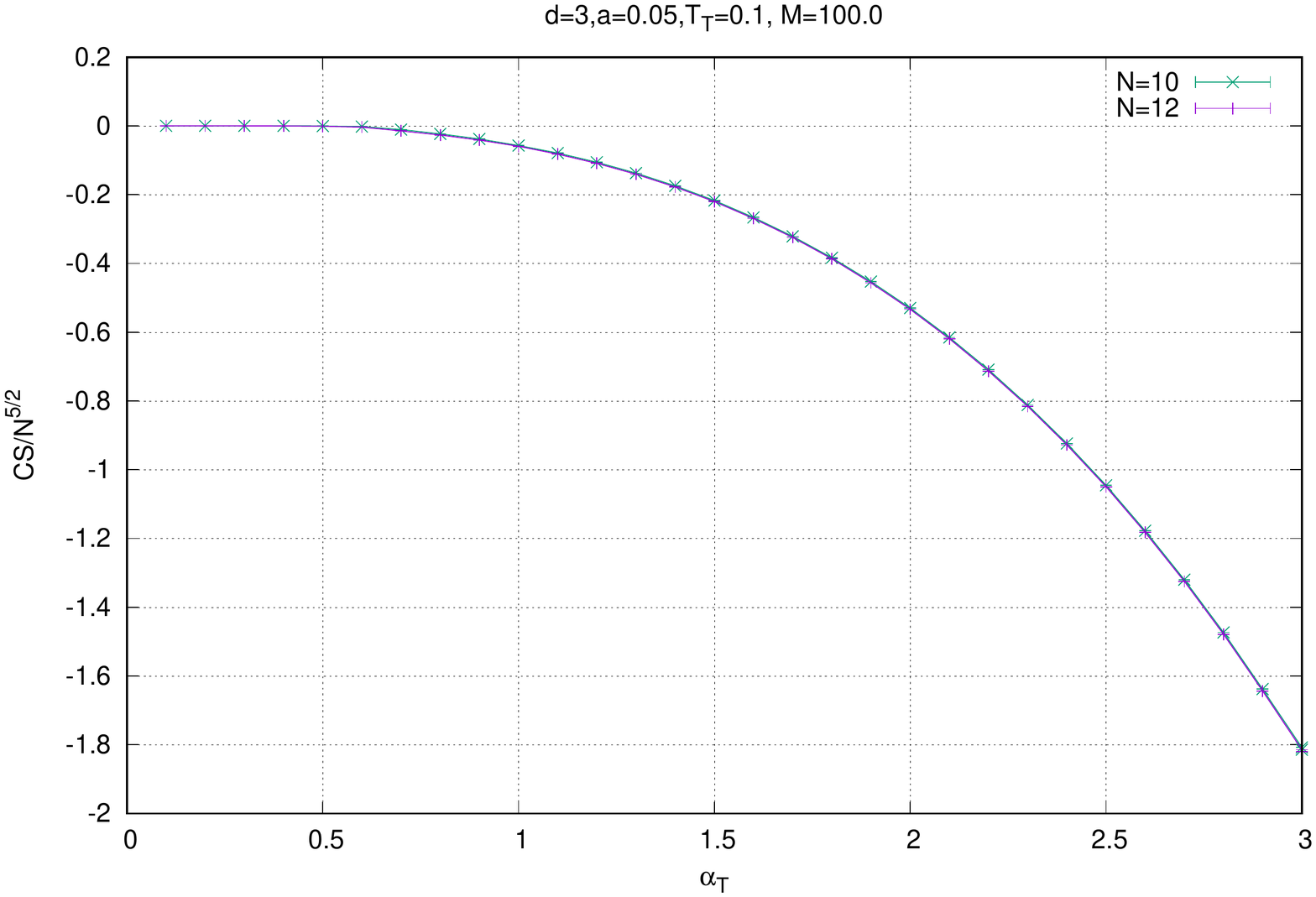}
  \includegraphics[width=10cm,angle=-0]{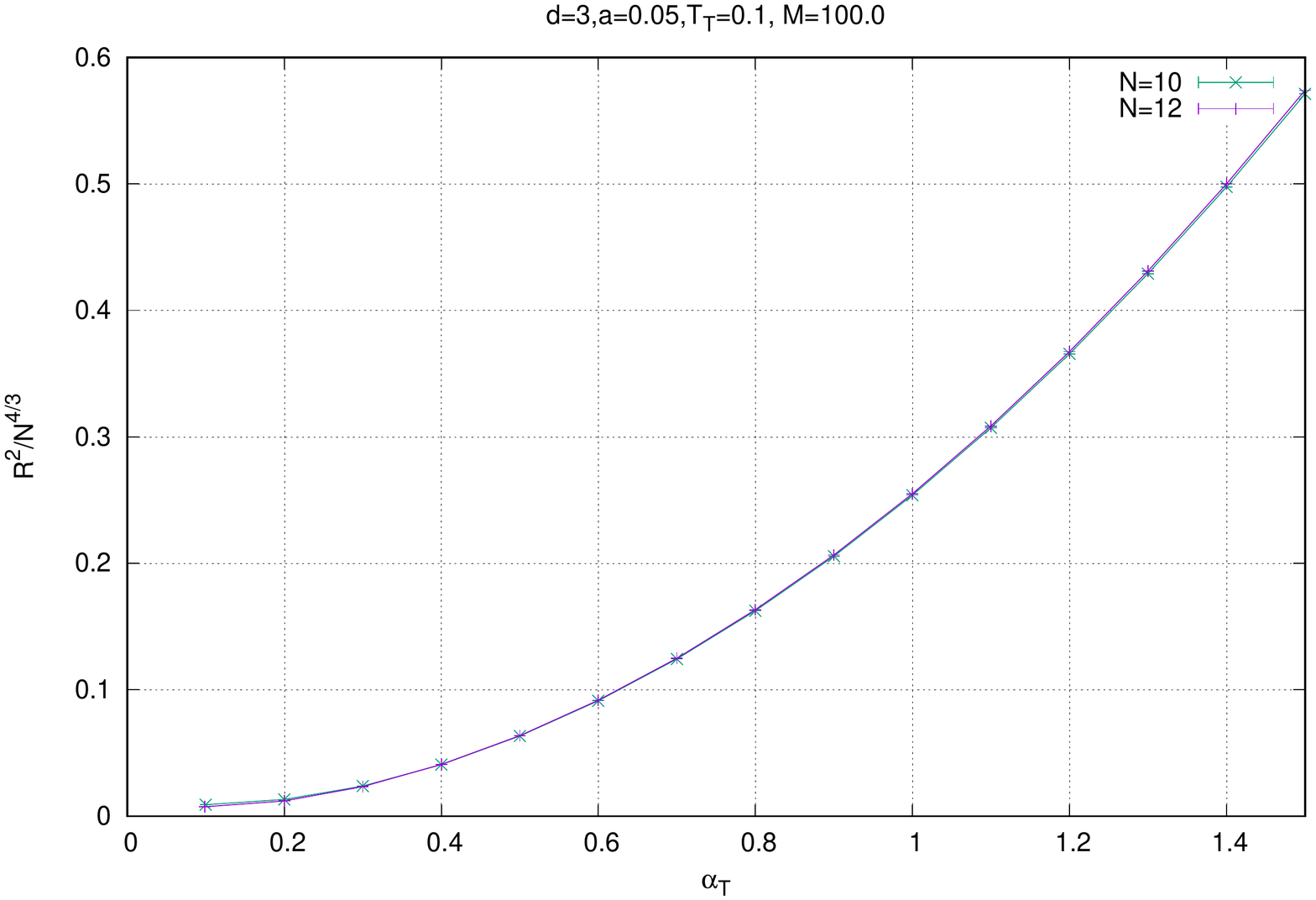}
  \includegraphics[width=10cm,angle=-0]{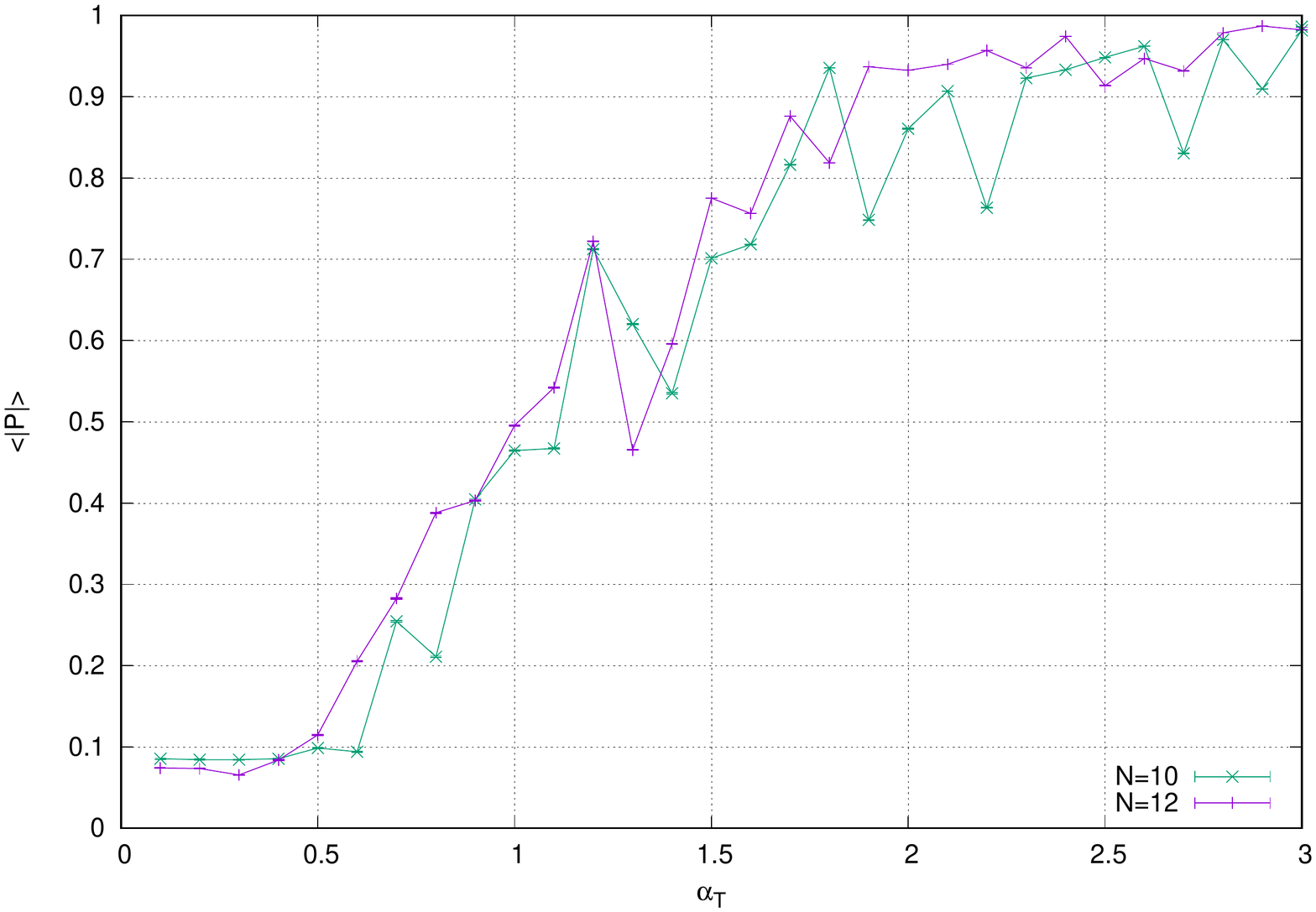}
\end{center}
\caption{The observables $\langle |P|\rangle$, $R^2$, ${\rm CS}$ for $\tilde{T}=0.1$. }\label{sample0v2}
\end{figure}

It is not difficult to check that the configuration (\ref{conf}) is a solution of the equation of motion (\ref{eom}) with the radius $\mu$ given explicitly by
\begin{eqnarray}
\mu=\frac{\alpha}{4M\hat{c}_2^{\prime}}\bigg[1\pm \sqrt{1+8M\hat{c}_2^{\prime}(2Mc_2-1)}\bigg]~,~\hat{c}_2^{\prime}=\frac{2}{N}\hat{c}_2.\label{conf1}
\end{eqnarray}
The values of the radius $R^2$ and of the Chern-Simons term ${\rm CS}$ in this configuration (in the large $M$ limit) are given by
\begin{eqnarray}
R^2=\frac{\tilde{\alpha}^2}{4}N^{4/3}~,~{\rm CS}=-\frac{\tilde{\alpha}^3}{6^{3/2}}N^{5/2}.\label{RCS}
\end{eqnarray}
The scaling of the gauge coupling constant $\alpha$ and of the temperature $T$ in the large $D$ model  $S_0+S_M$ is assumed to be identical to the scaling  found in the exact model (with the full Yang-Mills term instead of the harmonic oscillator term with $m=d^{1/3}$), viz \cite{Kawahara:2007nwF}
\begin{eqnarray}
  \tilde{\alpha}=\alpha N^{1/3}~,~\tilde{T}=\frac{T}{N^{2/3}}.\label{419}
\end{eqnarray}
The eigenvalue distributions of the matrices $D_a=\Phi_a(n)$ (for some fixed lattice site $n$) is measured in Monte Carlo simulations by the formula 
\begin{eqnarray}
  \rho(\lambda)&=&\frac{1}{N}\sum_{a=1}^N \langle\delta(\lambda-\lambda_a)\rangle.
\end{eqnarray}
But from equations (\ref{conf}) and (\ref{conf1}) we expect a continuum of eigenvalues centered around $\lambda=0$ together with two cuts at large values of $\lambda$ given explicitly by    
\begin{eqnarray}
\lambda_{\pm}  &=&\frac{\tilde{\alpha}}{2}N^{7/6}\frac{j_3}{\sqrt{2\hat{c_2}}}\nonumber\\
  &=&\pm \frac{\tilde{\alpha}}{2\sqrt{6}}N^{7/6}.
\end{eqnarray}
The scaling of the radius $R^2$, the Chern-Simons term ${\rm CS}$ and of the eigenvalues $\lambda_{\pm}$ are then given by $N^{4/3}$, $N^{5/2}$ and $N^{7/6}$ respectively which is confirmed by Monte Carlo simulations. For example,  this effect is shown for the temperature $\tilde{T}=0.1$ on figures (\ref{sample0v1}) (for the eigenvalues $\lambda_{\pm}$) and (\ref{sample0v2}) (for  $R^2$ and ${\rm CS}$). We also include in figure (\ref{sample0v2}) the result of the Polyakov line which shows that the baby fuzzy sphere exists in the deconfined phase ($\langle |P|\rangle$ is around $1$).

Indeed, it is seen from the Monte Carlo results of the eigenvalue distribution $\rho(\lambda)$ of the matrices $D_a$ that the transition from the fuzzy-sphere-like phase (the three-cut phase) to the Yang-Mills phase (the one-cut phase) at (say) the fixed temperature $\tilde{T}=0.1$ occurs at a value $\tilde{\alpha}_c=0.625\pm 0.125$ between $1$ and $0.5$. At this critical value $\tilde{\alpha}_c$ the Polyakov line switches from $0$ in the Yang-Mills phase to $1$ in the fuzzy-sphere-like phase.  On the other hand, the Chern-Simons term vanishes  whereas the radius becomes very small in the Yang-Mills phase. The critical line in the plane $\tilde{\alpha}-\tilde{T}$ can be constructed along these lines, i.e. by following the behavior of the eigenvalue distribution $\rho(\lambda)$ which we will do shortly.

The small eigenvalues  in (\ref{conf}) (which are fluctuating around $0$ with a degeneracy equal $N-2$) will play a greater role in the Yang-Mills phase (which is the phase encountered for smaller values of $\tilde{\alpha}$) and thus their behavior will be discussed next. 

\subsection{Yang-Mills phase and Wigner semi-circle law}

The eigenvalue distribution $\rho(\lambda)$ in the Yang-Mills phase becomes effectively a Wigner semi-circle law and it can be computed starting from the following two assumptions (confirmed by Monte Carlo simulations). First, the  Chern-Simons term vanishes in the Yang-Mills phase indicating the dominance of commuting and diagonal matrices. Second, the three commuting matrices $D_a$ are static and  by rotational invariance  their contributions are identical. The matrix model $S_0+S_M$ leads then to the effective potential (with $\beta_1=\beta.d$ and $\beta_2=\beta.d^2$)
\begin{eqnarray}
  V=N\beta_1(\frac{m^2}{2}-2Mc_2\varphi)TrM^2+M\beta_2(TrM^2)^2.
\end{eqnarray}
The saddle point equation of this matrix model is identical to the saddle point equation of the quadratic matrix model $BTrM^2$ with the mass $B$ given by
\begin{eqnarray}
B=N\beta_1(\frac{m^2}{2}-2Mc_2\varphi)+2M\beta_2Na_2=NB_0+NB_1a_2.\label{Bmass}
\end{eqnarray}
The difference with the usual quadratic matrix model is the fact that the parameter $B$ depends on the second moment $a_2=TrM^2/N$. The solution of 
 the matrix model $BTrM^2$ is known to be given by the Wigner semi-circle law
\begin{eqnarray}
  \rho(\lambda)=\frac{2}{\pi a^2}\sqrt{a^2-\lambda^2}~,~a^2=\frac{2N}{B}.
\end{eqnarray}
From this solution we compute the second moment $a_2=\int_{-a}^{+a} d\lambda \rho(\lambda)\lambda^2=a^2/4=N/2B$. This leads to the consistency condition $Ba_2=N/2$ or equivalently to the quadratic equation $B_1a_2^2+B_0a_2-1/2=0$  which admits the solution
\begin{eqnarray}
a_2=\frac{\sqrt{B_0^2+2B_1}-B_0}{2B_1}.
\end{eqnarray}
By substituting in the mass parameter $B$ we obtain the solution
\begin{eqnarray}
B=\frac{N}{2}(\sqrt{B_0^2+2B_1}+B_0).
\end{eqnarray}
Hence, the radius of the Wigner semi-circle is given by
\begin{eqnarray}
a^2=2\frac{\sqrt{B_0^2+2B_1}-B_0}{B_1}.\label{law}
\end{eqnarray}
Here there are two quite different behaviors. First, for $\tilde{\alpha}\neq 0$ we get for large $M$ the solution (corresponding to commuting matrices)
\begin{eqnarray}
a^2=\frac{4\beta_1c_2\varphi}{\beta_2}+\frac{2}{B_1}(\frac{\beta_2}{\beta_1c_2\varphi}-\beta_1m^2)+O(\frac{1}{M^2})\Rightarrow a=\frac{\tilde{\alpha}N^{2/3}}{\sqrt{d}}+O(\frac{1}{M}).\label{law1}
\end{eqnarray}
This leading value of the radius is independent of $M$ and $\tilde{T}$, it depends linearly on $\tilde{\alpha}$ and scales slowly with $N$ as $N^{2/3}$.

\begin{figure}[htbp]
\begin{center}
  \includegraphics[width=10cm,angle=-0]{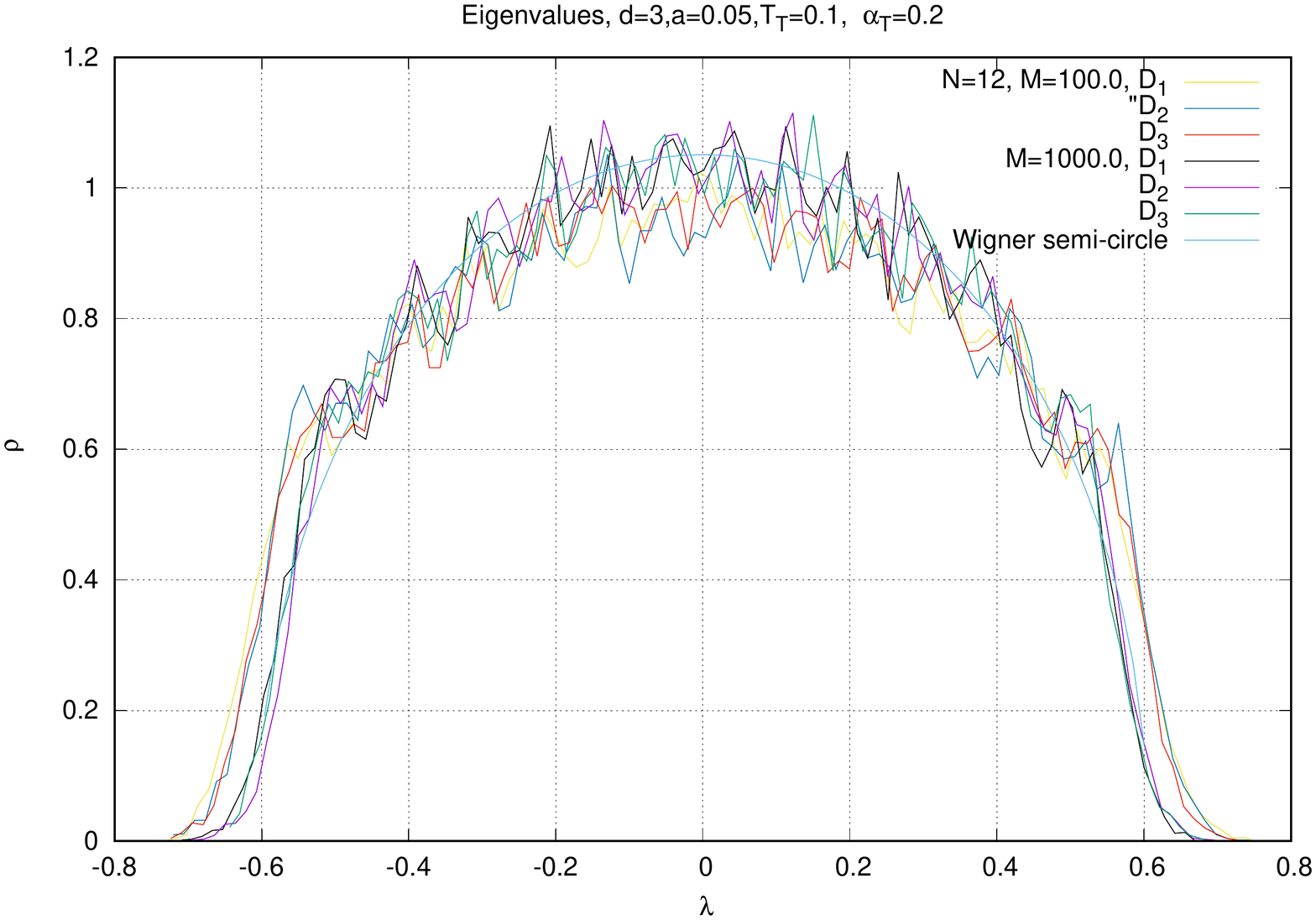}
  \includegraphics[width=10cm,angle=-0]{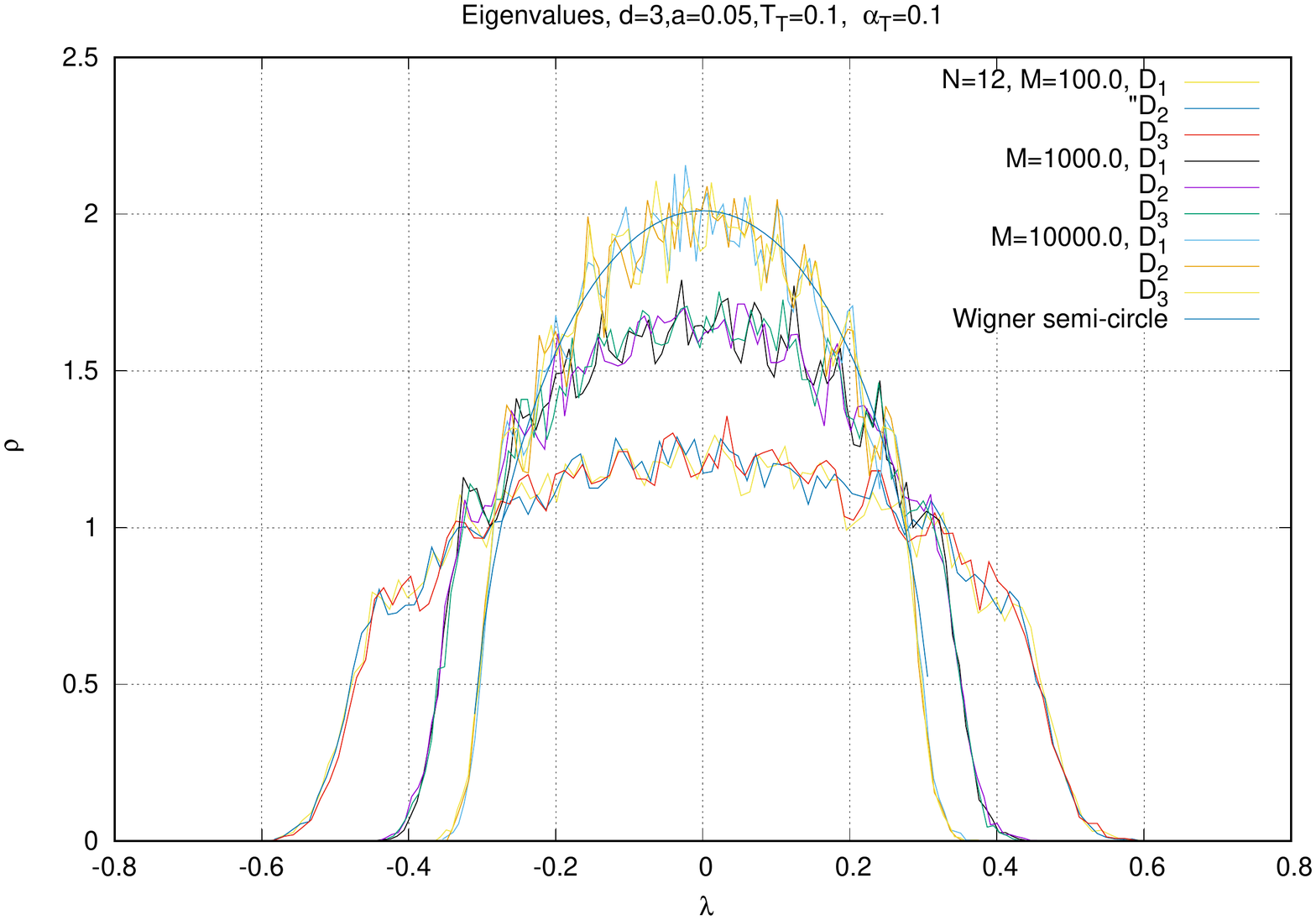}
\end{center}
\caption{The Wigner's semi-circle law for $\tilde{T}=0.1$. }\label{sample0v3}
\end{figure}

However, for $\tilde{\alpha}=0$ the behavior for large $M$ is found to be distinctly different (corresponding to the minimum $\Phi_a=0$ exactly) given by the law 
\begin{eqnarray}
a=(\frac{4}{M\beta d^2})^{1/4}.\label{law2}
\end{eqnarray}
Both behaviors (\ref{law1}) and (\ref{law2}) are confirmed in Monte Carlo simulations. See the two graphs on figure (\ref{sample0v3}) for $\tilde{\alpha}=0.2$ and $\tilde{\alpha}=0.1$ respectively (with $\tilde{T}=0.1$ and $N=12$).

For later purposes, it is also important to keep track of the leading $1/M$ correction which contains the $T-$dependence of the result (\ref{law1}). We have then the radius  of the Wigner semi-circle law 
\begin{eqnarray}
a^2=\frac{4c_2\alpha^2}{d}+\frac{1}{Md}(\frac{T}{c_2\alpha^2}-2\alpha^2)+O(\frac{1}{M^2}).\label{law3}
\end{eqnarray}
This behavior is also confirmed in Monte Carlo simulations for high temperatures (see next section).

The radius $R^2$ can then be calculated by the formula
\begin{eqnarray}
  R^2&=&da_2\nonumber\\
  &=&\frac{d}{4}a^2\nonumber\\
  &=&c_2\alpha^2+\frac{1}{4M}(\frac{T}{c_2\alpha^2}-2\alpha^2)+O(\frac{1}{M^2}).\label{beauty}
  \label{eom1}
\end{eqnarray}
Thus, the leading behavior agrees with the result in the baby sphere phase given by the first equation in (\ref{RCS}) which indicates that this observable is continuous across the critical point. Hence, the  leading $1/M$ correction must also be continuous.

By approximating  $Tr\Phi_a^2/N$ by $R^2$ and then using equation (\ref{eom1}) we can rewrite the equation of motion (\ref{eom}) for static configurations as
  \begin{eqnarray}
    -i\alpha\epsilon_{abc}[\Phi_b,\Phi_c]&=&2(\alpha^2+2MR^2-2Mc_2\alpha^2)\Phi_a\nonumber\\
    &=&\frac{T}{c_2\alpha^2}\Phi_a.
\end{eqnarray}
  This lead to the fuzzy sphere algebra given by
  \begin{eqnarray}
[\Phi_a,\Phi_b]=\frac{i}{N^{1/3}}\frac{2\tilde{T}}{\tilde{\alpha}^3}\epsilon_{abc}\Phi_c.
  \end{eqnarray}
  Naturally this is vanishingly  small in the large $N$ limit which shows explicitly how the geometry decays in the Yang-Mills phase.

\subsection{The critical boundary and temperature dependence}

In order to study the temperature dependence of the theory in Monte Carlo simulations we  need to vary the temperature $T=1/\beta$ which is tied to the lattice spacing $a$ by the relation $\beta=\Lambda.a$. In the previous two sections we have studied the low temperature region  (with $T=0.1$) with a value of the lattice spacing given by $a=0.05$ which we believe was sufficient to uncover the true continuum limit at low $T$. But as we start increasing the temperature we have to start changing  the lattice spacing $a$ and/or the number of lattice sites $\Lambda$ for each temperature $T$ in such a way as to keep the finite lattice spacing effects at bay and as a consequence reach the true continuum limit. This can be become very involved at high temperatures since the regularization term (\ref{ac1}) becomes suppressed as we start increasing the temperature and as a consequence  thermalization becomes more time (Monte Carlo time) consuming as we will now show.

To see that the regularization term (\ref{ac1}) becomes suppressed at high temperatures we go back to the Yang-Mills quantum mechanics $S_0+S_M$ given by (\ref{ac0})+(\ref{ac1}) together with (\ref{ac2}) and rewrite it in the form
\begin{eqnarray}
S_1=N\int_0^{\beta}dt Tr\bigg[\frac{1}{2}(D_t\Phi_a)^2+\alpha^2\Phi_a^2+\frac{2i\alpha}{3}\epsilon_{abc}\Phi_a\Phi_b\Phi_c\bigg]+M\int_0^{\beta}dt (Tr\Phi_a^2-\alpha^2 c_2 N)^2.
\end{eqnarray}
Under the rescaling $\hat{t}=t/\rho$, $\hat{\beta}=\beta/\rho$, $\hat{\Phi}_a=\Phi_a/\sqrt{\rho}$, $\hat{A}=A\rho $, $\hat{\alpha}=\alpha \rho^{5/2}$ and $\hat{M}=M\rho^3$ we obtain 
\begin{eqnarray}
S_1=N\int_0^{\hat{\beta}}d\hat{t} Tr\bigg[\frac{1}{2}(\hat{D}_{\hat{t}}\hat{\Phi}_a)^2+\frac{\hat{\alpha}^2}{\rho^3}\hat{\Phi}_a^2+\frac{2i\hat{\alpha}}{3}\epsilon_{abc}\hat{\Phi}_a\hat{\Phi}_b\hat{\Phi}_c\bigg]+\hat{M}\int_0^{\hat{\beta}}d\hat{t} (Tr\hat{\Phi}_a^2-\frac{\hat{\alpha}^2}{\rho^6} c_2 N)^2.
\end{eqnarray}
The lattice spacing is now given by $\hat{a}=\hat{\beta}/\Lambda$. An optimal value of the rescaled temperature $\hat{T}$ (between low and high temperatures as observed in Monte Carlo simulations) is  $\hat{\beta}=1$. We choose then  the scale $\rho=1/T$ leading to 
the following rescalings
\begin{eqnarray}
\hat{m}^2=2\hat{\alpha}^2T^3~,~\hat{\alpha}=\alpha/T^{5/2}~,~\hat{M}=\frac{M}{T^3}~,~\hat{\varphi}=\hat{\alpha}^2T^6.
\end{eqnarray}
We can see immediately that the regularization mass $\hat{M}$ is suppressed at high temperatures $T$ as $1/T^3$.  The condition (\ref{constraint0}) corresponding to the regularization term (\ref{ac1}) will then be supplemented at high temperatures with the usual tracelessness condition $Tr \Phi_a(n)=0$. 

\begin{figure}[htbp]
\begin{center}
  \includegraphics[width=10cm,angle=-0]{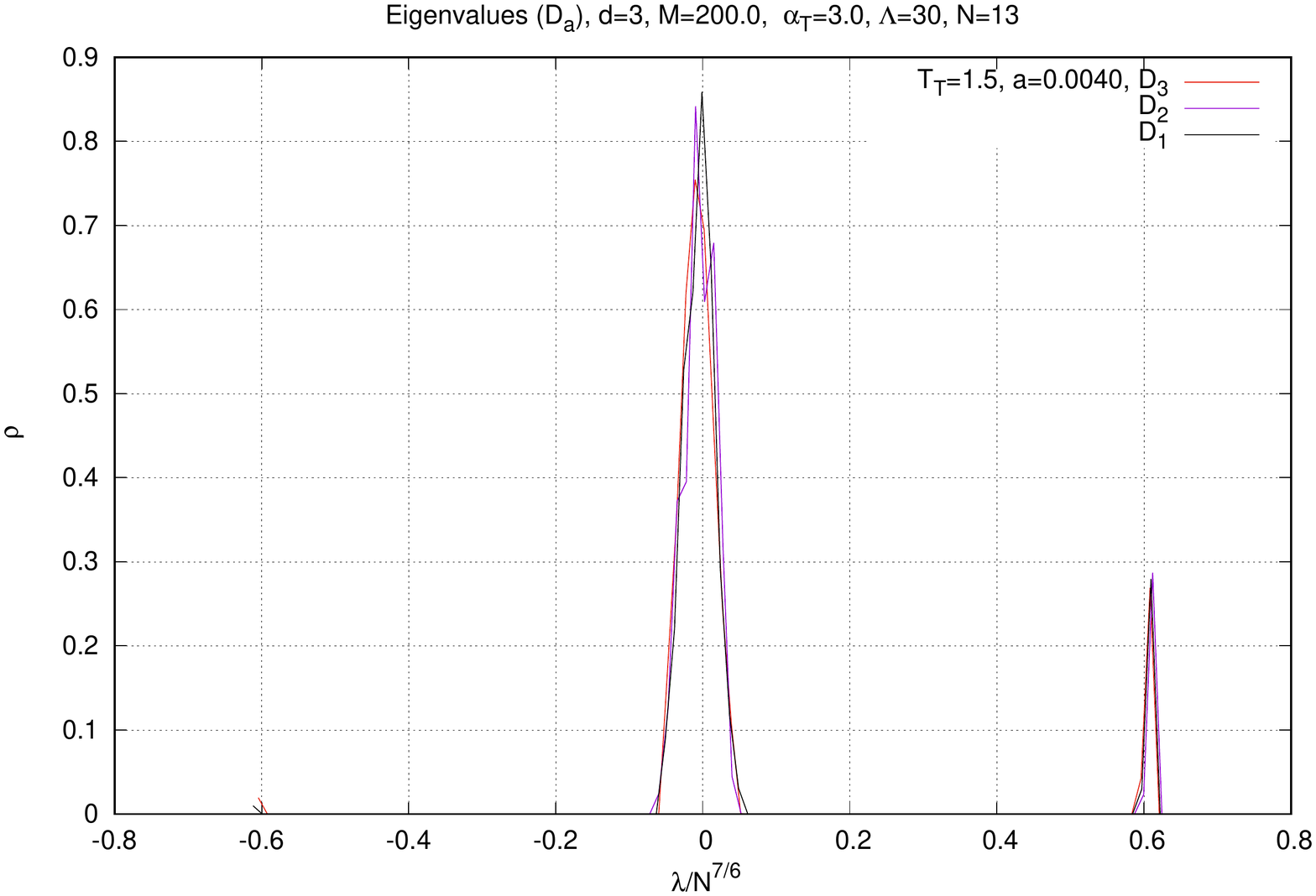}
  \includegraphics[width=10cm,angle=-0]{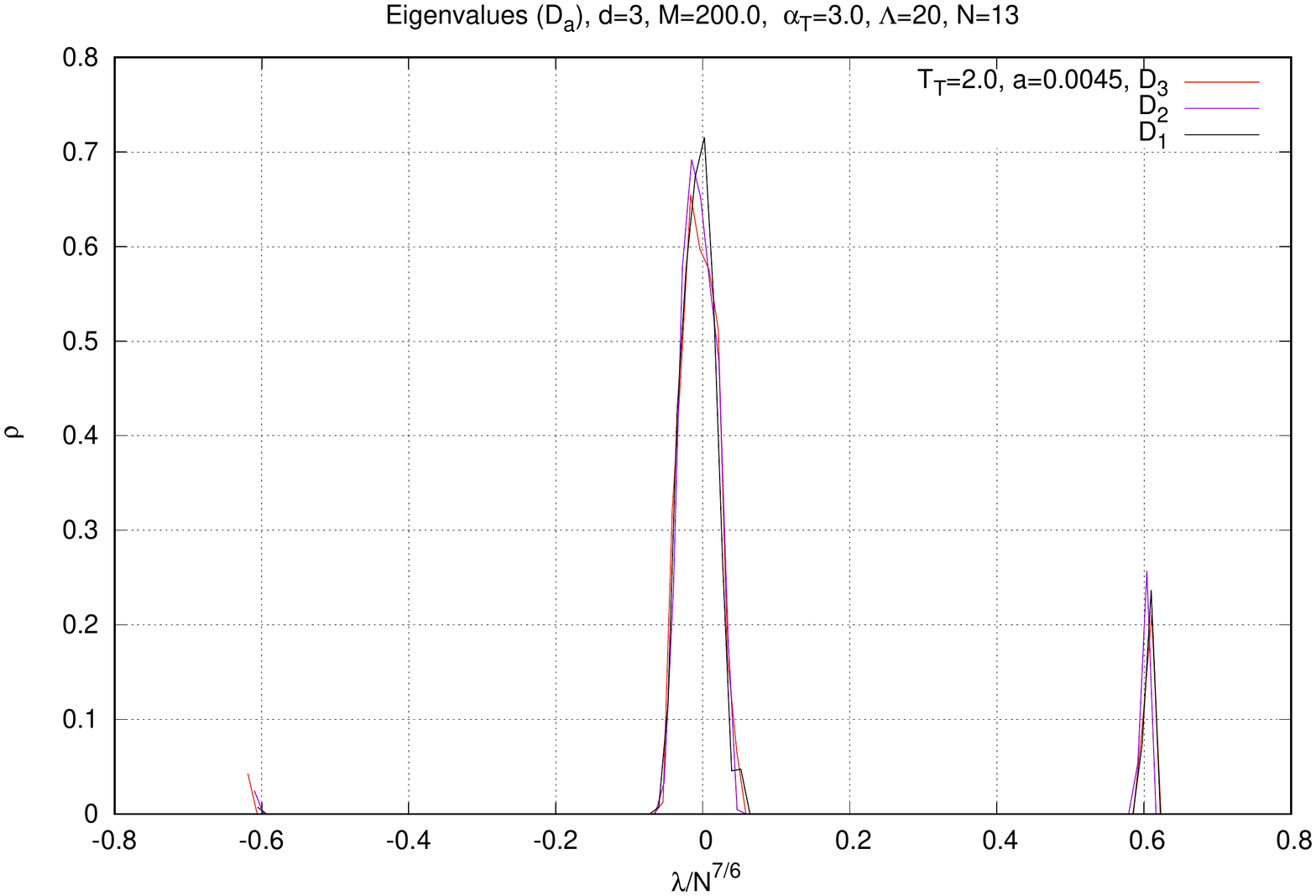}
  \includegraphics[width=10cm,angle=-0]{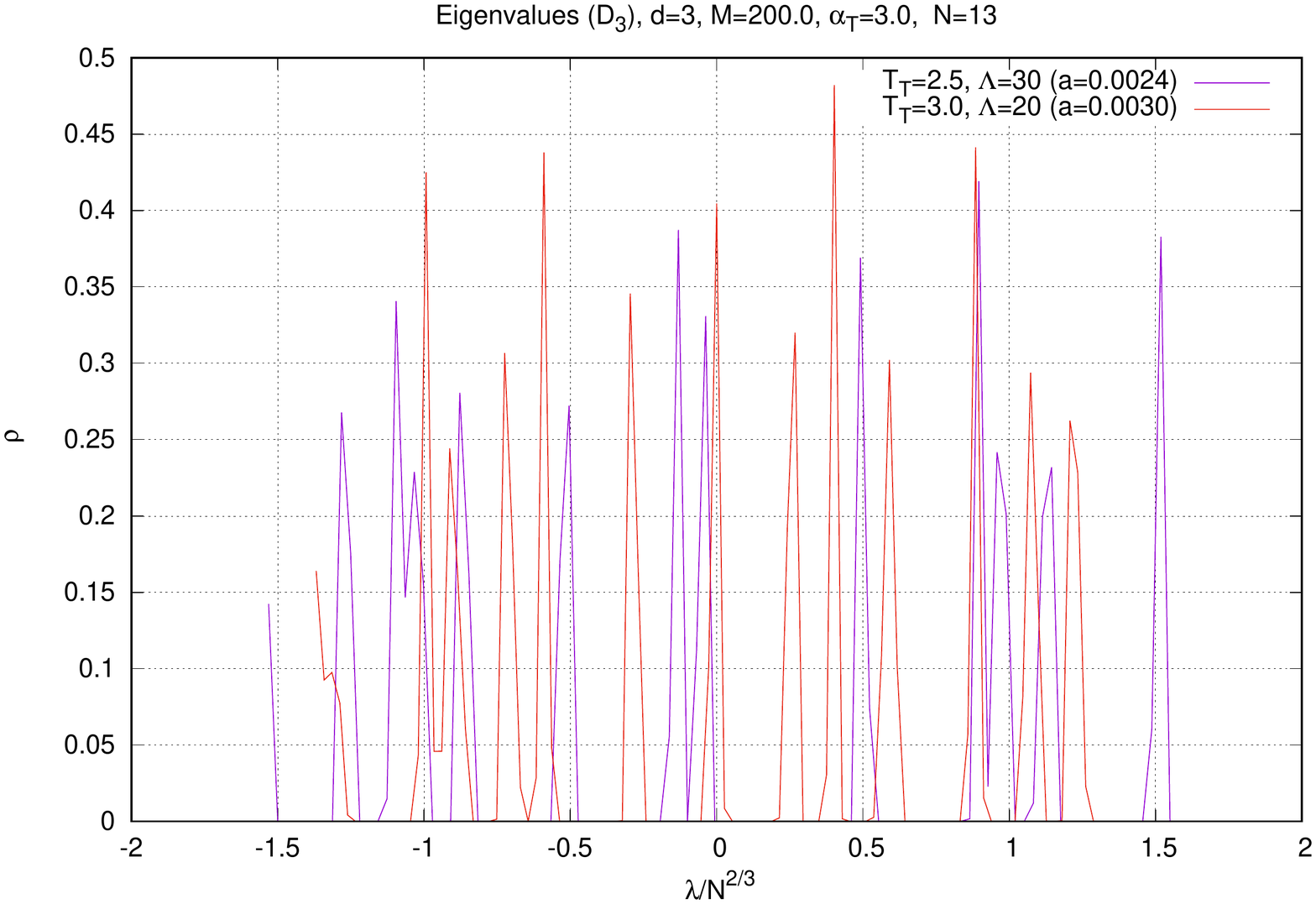}
\end{center}
\caption{The  eigenvalue distribution $\rho(\lambda)$ of the matrices $D_a$  for $\tilde{\alpha}=3$ and $N=13$.}\label{sample1v1}
\end{figure}

\begin{figure}[htbp]
\begin{center}
  \includegraphics[width=10cm,angle=-0]{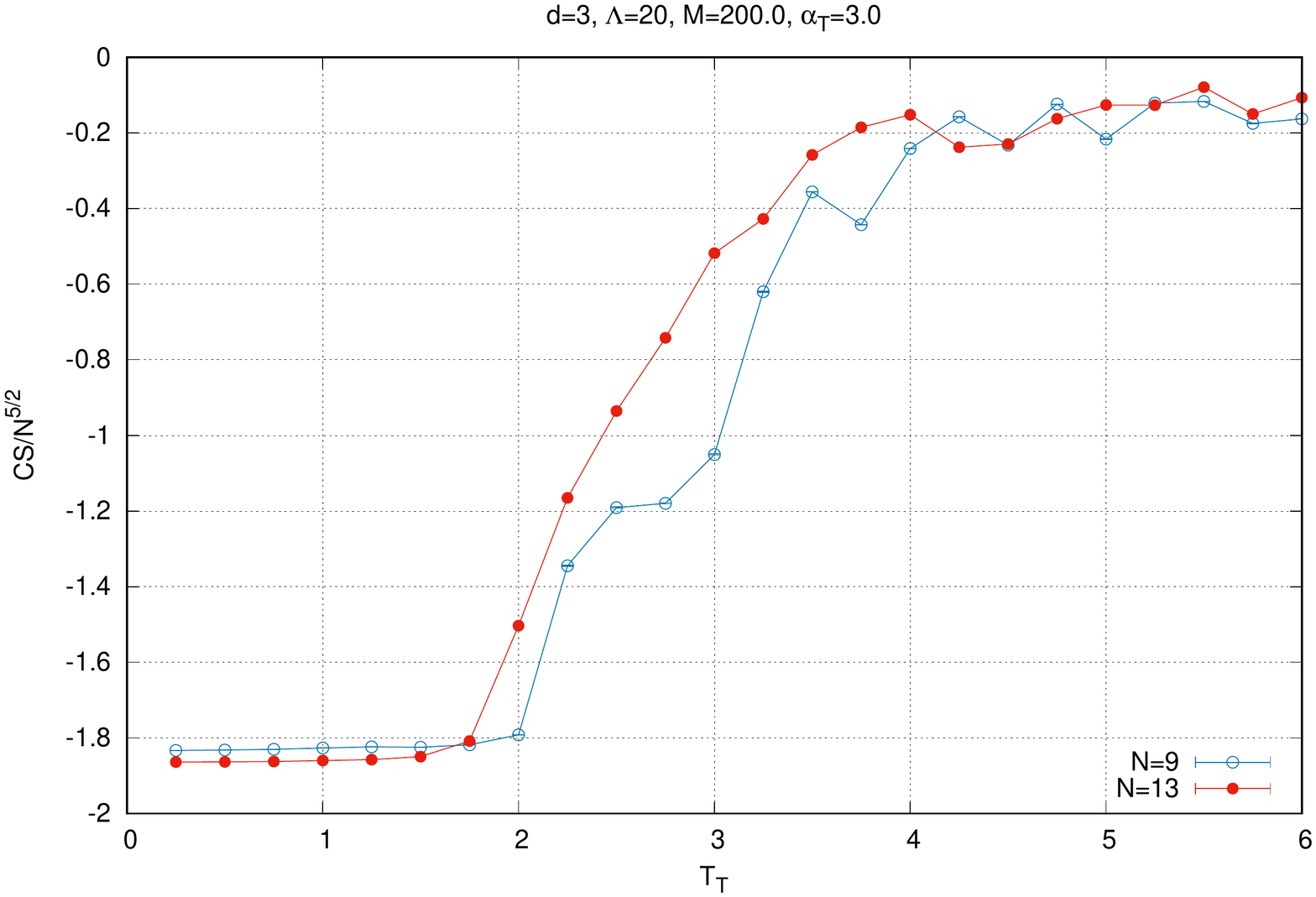}
  \includegraphics[width=10cm,angle=-0]{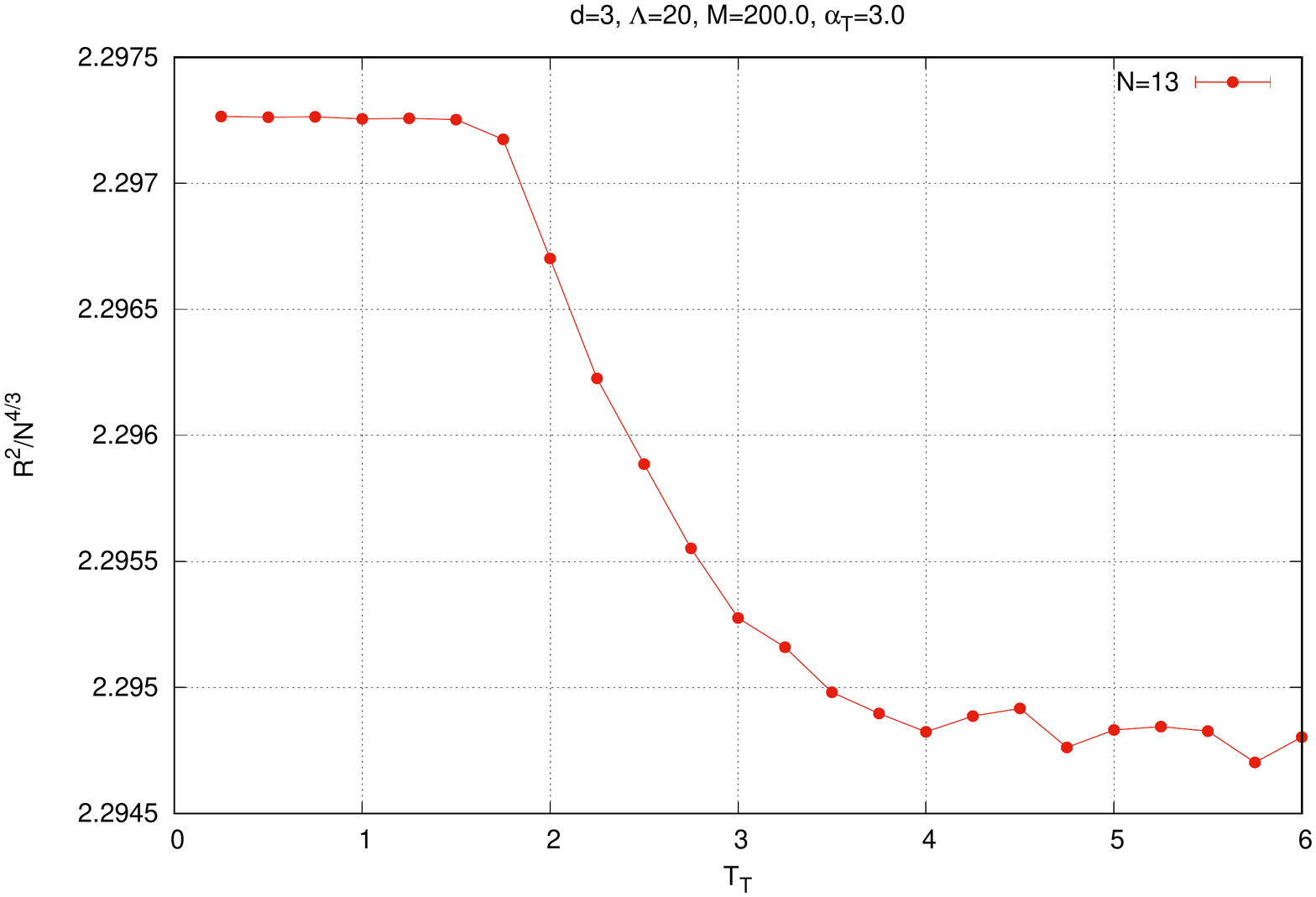}
  \includegraphics[width=10cm,angle=-0]{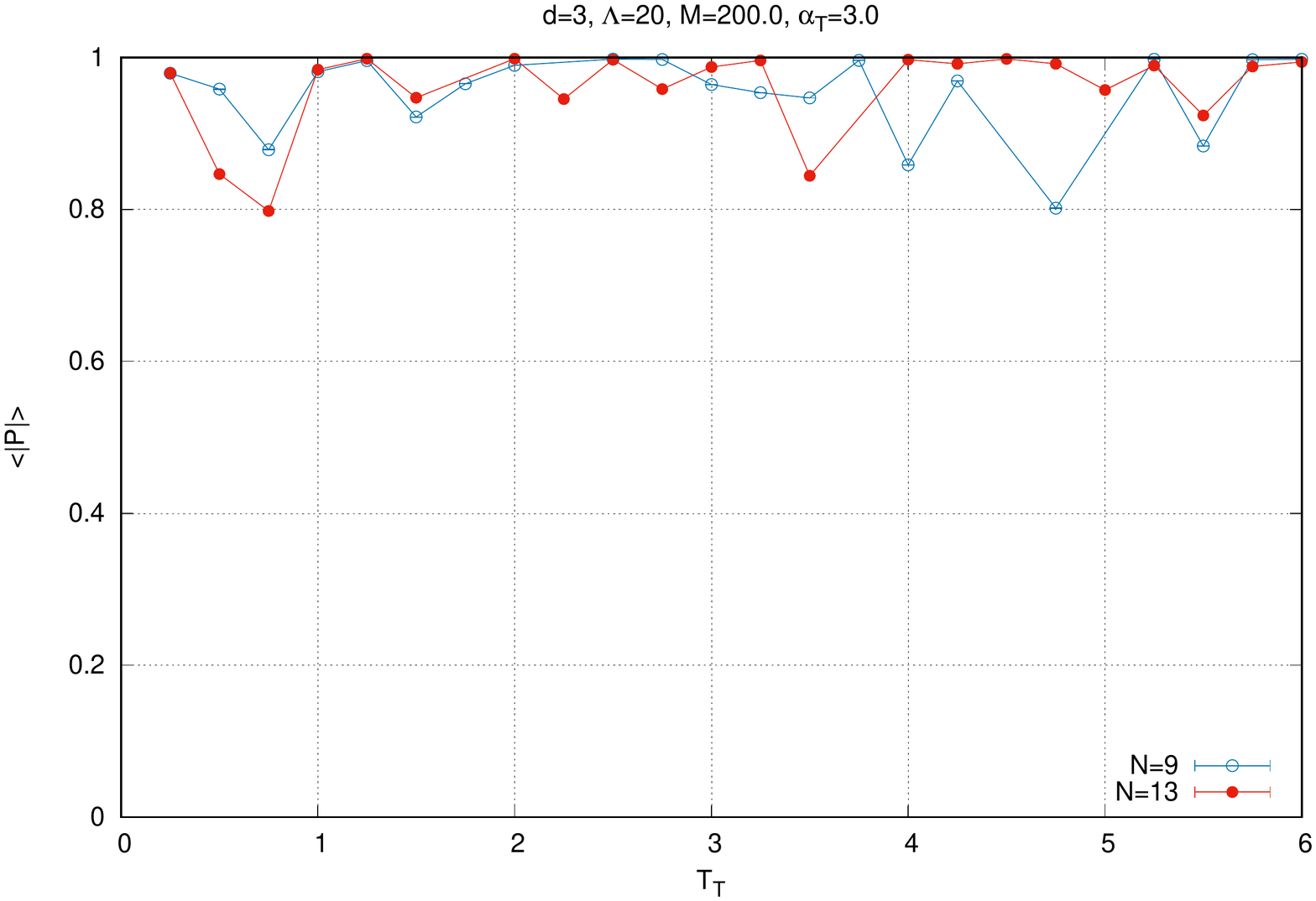}
\end{center}
\caption{The observables $\langle |P|\rangle$, $R^2$, ${\rm CS}$ for $\tilde{\alpha}=3$  and $\Lambda=20$. }\label{sample2}
\end{figure}

We start our systematic study of the temperature-dependence of this model by considering the behavior of the various observables at large values of the gauge coupling constant $\tilde{\alpha}$. In other words, we cross the phase diagram $\tilde{T}-\tilde{\alpha}$ horizontally. For example, we consider   in the two samples (\ref{sample1v1}) and (\ref{sample2}) the behavior at $\tilde{\alpha}=3$ of the eigenvalue distribution $\rho(\lambda)$ and the observables  $\langle |P|\rangle$, $R^2$, ${\rm CS}$ (given by equations (\ref{obs1}), (\ref{obs2}) and (\ref{obs3})) respectively \footnote{In this first sample (\ref{sample1v1}) we run the Metropolis algorithm for $2^{14}+2^{13}$ steps (except at the transition points where we thermalize for  $2^{15}+2^{13}$ steps) where the first $2^{14}(2^{15})$ steps are discarded whereas histograms and averages are computed using the last $2^{13}$ steps. We choose two lattices $\Lambda=20$ and $\Lambda=30$. In the second sample  (\ref{sample2}) we run the Metropolis algorithm for $2^{14}+2^{13}$ steps and use only the lattice $\Lambda=20$}.

The continuum limit $\Lambda\longrightarrow \infty$ and the commutative limit $N\longrightarrow \infty$ of the eigenvalue distribution $\rho(\lambda)$ of the matrices $D_a=\Phi_a(n)$ (where $n$ is taken to be the first point in the lattice) for $\tilde{\alpha}=3$  across the critical boundary are shown in figure (\ref{sample1v1}) (from here on we also choose the mass value $M=200$).

The transition point $\tilde{T}_*$ is seen as the value of the temperature where various observables suffer a jump as we cross from the baby fuzzy
sphere phase into the Yang-Mills phase (from our data we think this is a discontinuous transition in accordance with \cite{Kawahara:2007nwF}). We observe that the critical point $\tilde{T}_*$ depends on the matrix size $N$ (here we take $13$) and the result is consistent with the one-loop estimation (\ref{oneloop}) (see next section). On the lattice $\Lambda=20$ we get the value $2.5\pm 0.5~(N=13)$ whereas on the lattice $\Lambda=30$ we get the value $1.75\pm 0.25~(N=13)$. We should extrapolate these values to $\Lambda\longrightarrow \infty$ in order to obtain the continuum values or we can simply take as an approximation the estimation obtained with the larger lattice $\Lambda=30$ and compare it with the one-loop result  (\ref{oneloop}).

Thus, the transition temperature $\tilde{T}_*$ marks the switch between the baby fuzzy sphere phase and the Yang-Mills phase. The baby fuzzy sphere phase is, as we have discussed before,  characterized by  the three-cut solution

\begin{eqnarray}
\Phi_a=\bigg(\begin{array}{cc}
      \mu (J_a)_{2\times 2} & 0 \\
      0 & 0
      \end{array}\bigg)~,~\mu= \pm \frac{\alpha N}{2\sqrt{\hat{c_2}^{\prime}}}~,~M\longrightarrow\infty.
\end{eqnarray}
The Yang-Mills is only characterized by a Wigner semi-circle law in the limit of small values of the gauge coupling constant $\tilde{\alpha}\longrightarrow 0$. In the generic case, the  Yang-Mills phase is characterized by commuting matrices as shown by the vanishing of the Chern-Simons term in the first graph of figure  (\ref{sample2}). At the transition point it is observed that both the Chern-Simons and the radius jump ($|{\rm CS}|$ decreases and $R^2$ increases)  as we cross the critical boundary from the baby fuzzy sphere solution (lower values of $\tilde{T}$) to the Yang-Mills phase (higher values of the temperatures). It is also seen from the associated result of the Polyakov line that the baby fuzzy sphere exists always in the deconfined phase ($\langle |P|\rangle$ is around $1$).

The Yang-Mills phase is thus dominated by commuting matrices which for small values of the gauge coupling constant $\tilde{\alpha}$ are characterized by the Wigner semi-circle law whereas for large values of the gauge coupling constant (like our current case) are seen to be given by a multi-cut distribution which can be approximated by a uniform distribution \cite{OConnor:2012vwcF} (or the other way around) of the generic form 
\begin{eqnarray}
  \rho(\lambda)&=&\frac{1}{N}\sum_{i=1}^N  \delta(\lambda-\lambda_i)\Leftrightarrow\rho(\lambda)=\frac{1}{2a}.
\end{eqnarray}
Here $a$ is the radius of the uniform distribution (determined below).

However, the matrices $D_a$ must still satisfy the constraint (\ref{constraint0}). By rotational invariance we can make the approximation $TrD_a^2=d TrD_1^2$ and rewrite  (\ref{constraint0}) in terms of the eigenvalues $\lambda_i$ of $D_1$ as follows (we are also assuming $M\longrightarrow \infty$)
\begin{eqnarray}
d(\lambda_1^2+...+\lambda_N^2)=a^2~,~a=\frac{\tilde{\alpha}}{2}N^{2/3}.
\end{eqnarray}
By undoing the approximation $TrD_a^2=d TrD_1^2$ we obtain the more precise result
\begin{eqnarray}
\boxed{(\lambda_1^2+...+\lambda_N^2)+(\eta_1^2+...+\eta_N^2)+(\xi_1^2+...+\xi_N^2)=a^2~,~a=\frac{\tilde{\alpha}}{2}N^{2/3}.}
\end{eqnarray}
$\{\eta_i\}$ and $\{\xi_i\}$ are the eigenvalues of $D_2$ and $D_3$ respectively.

The eigenvalues are then uniformly distributed inside a solid ball as predicted in  \cite{OConnor:2012vwcF} with radius $a$ which scales as $N^{2/3}$. This is confirmed in the relevant graphs in figure  (\ref{sample1v1}) which depict the eigenvalue distribution $\rho(\lambda)$ immediately as we go through the critical boundary into the Yang-Mills phase. The maximum eigenvalue is also seen to be given indeed by the value $a/N^{2/3}=\tilde{\alpha}/2$. 

Next we study the behavior  of the various observables at fixed values of the temperature, i.e. we cross the phase diagram $\tilde{T}-\tilde{\alpha}$ vertically. The main difference between low and high temperatures is the fact that the transition from the baby fuzzy sphere phase configuration (at larger values of the gauge coupling constant $\tilde{\alpha}$) to the Wigner semi-circle law (at very small values of $\tilde{\alpha}$) goes for high temperatures through the phase of the commuting matrices.

\begin{figure}[htbp]
\begin{center}
  \includegraphics[width=15cm,angle=-0]{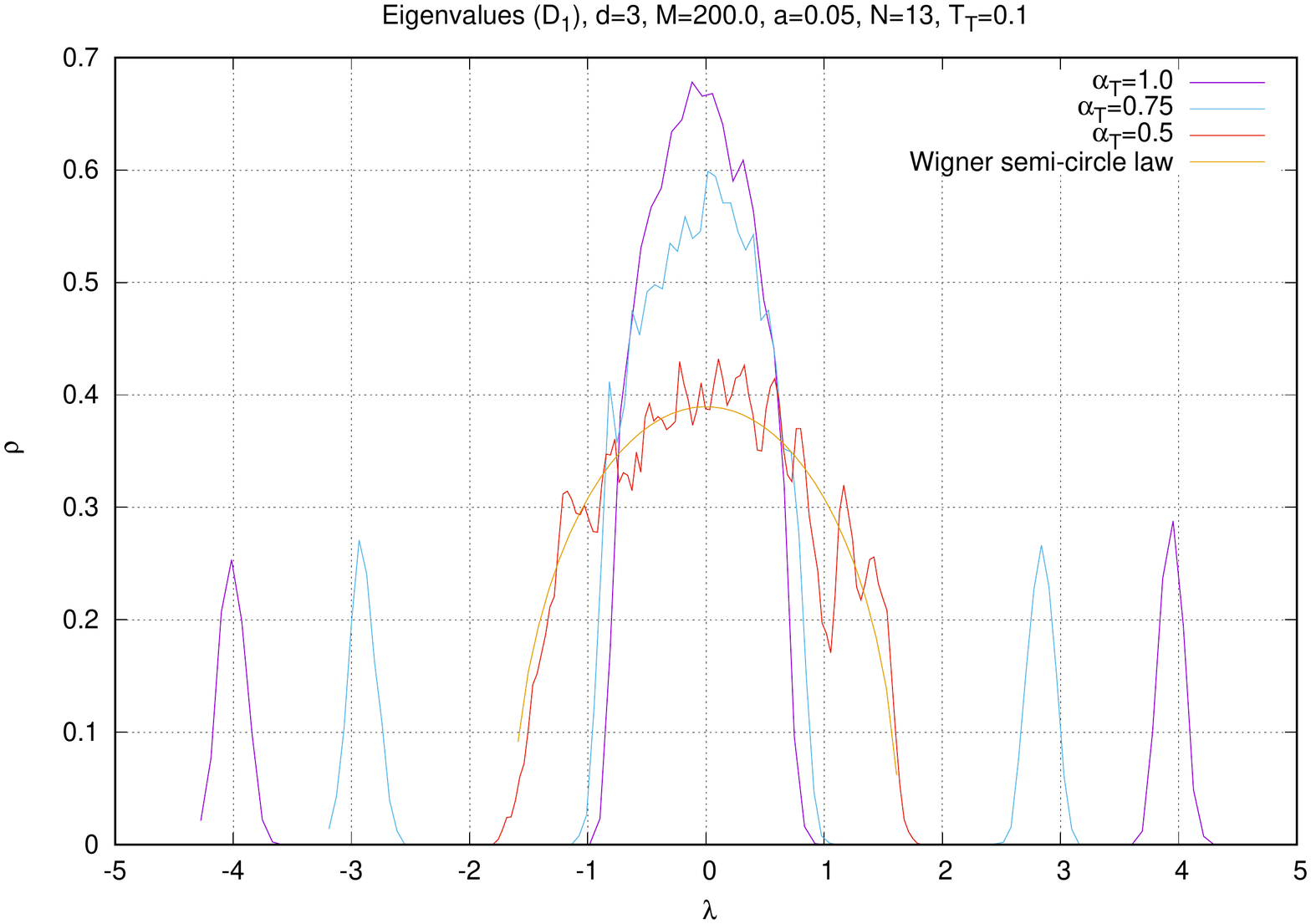}
\end{center}
\caption{The  eigenvalue distribution $\rho(\lambda)$ of the matrices $D_a$  for $\tilde{T}=0.1$, $N=13$ and $a=0.05$.}\label{sample5}
\end{figure}

But at low temperatures the transition from the baby fuzzy sphere phase to the Wigner semi-circle law is immediate which is simply explained by the fact that the critical boundary at low temperatures is located at very small values of $\tilde{\alpha}$. See the eigenvalue distribution for $\tilde{T}=0.1$ and $N=13$ in figure (\ref{sample5}). We measure for both $\tilde{T}=0.1$ on a lattice $a=0.05$ and $\tilde{T}=0.5$ on a lattice $\Lambda=30$ the critical value $\tilde{\alpha}=0.625\pm 0.125$ \footnote{At low temperatures thermalization is fast. In most cases we run the Metropolis algorithm for $2^{14}+2^{13}$ steps where  the first $2^{14}$ steps  are discarded whereas histograms and averages are computed using the last $2^{13}$ steps.}.  

As we increase $\tilde{T}$ the Yang-Mills phase opens up  and the phase of commuting matrices with a uniform distribution emerges separating the baby fuzzy sphere phase and the Wigner semi-circle law (in some sense we have in the phase diagram really three phases not two connected to emergent geometry).

\begin{figure}[htbp]
\begin{center}
  \includegraphics[width=10cm,angle=-0]{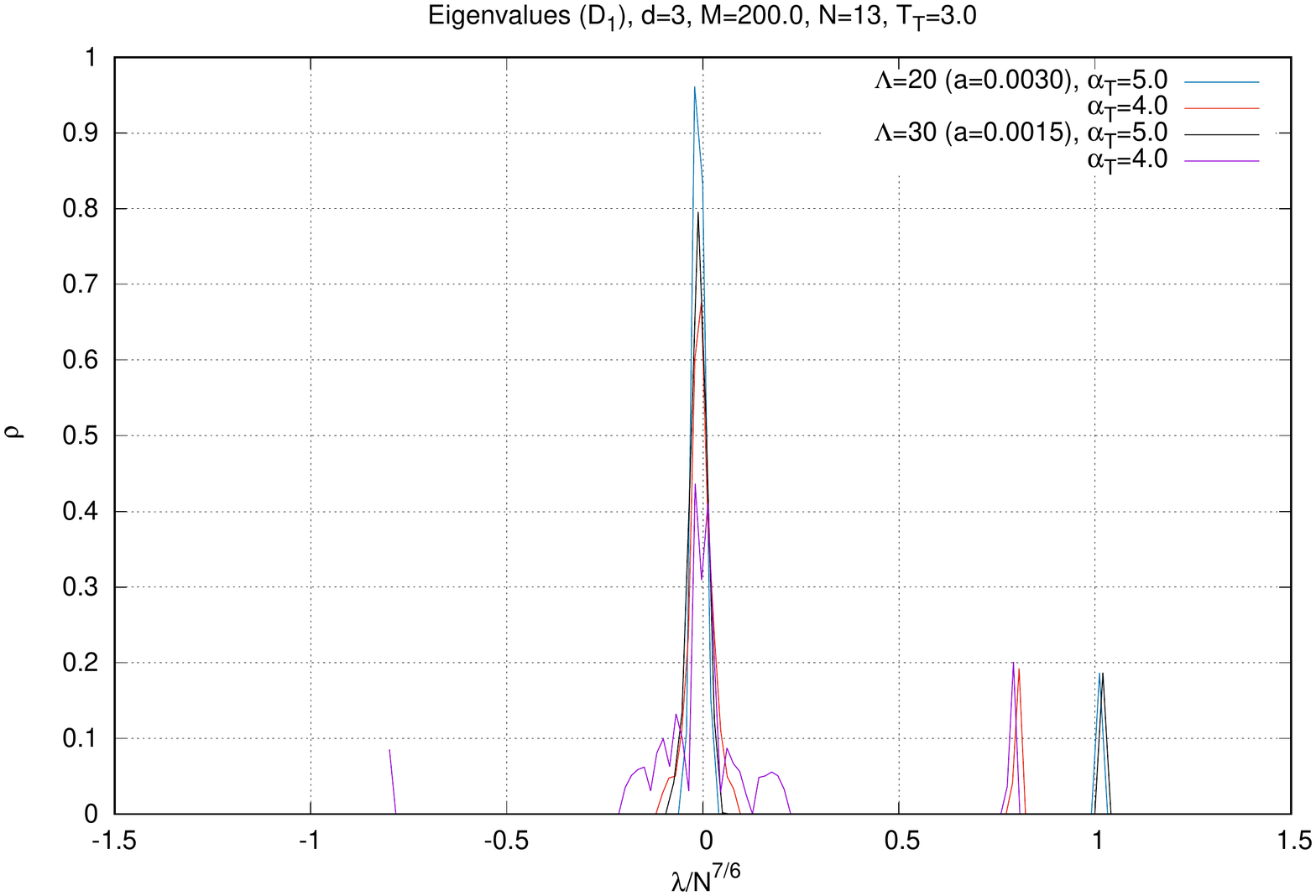}
  \includegraphics[width=10cm,angle=-0]{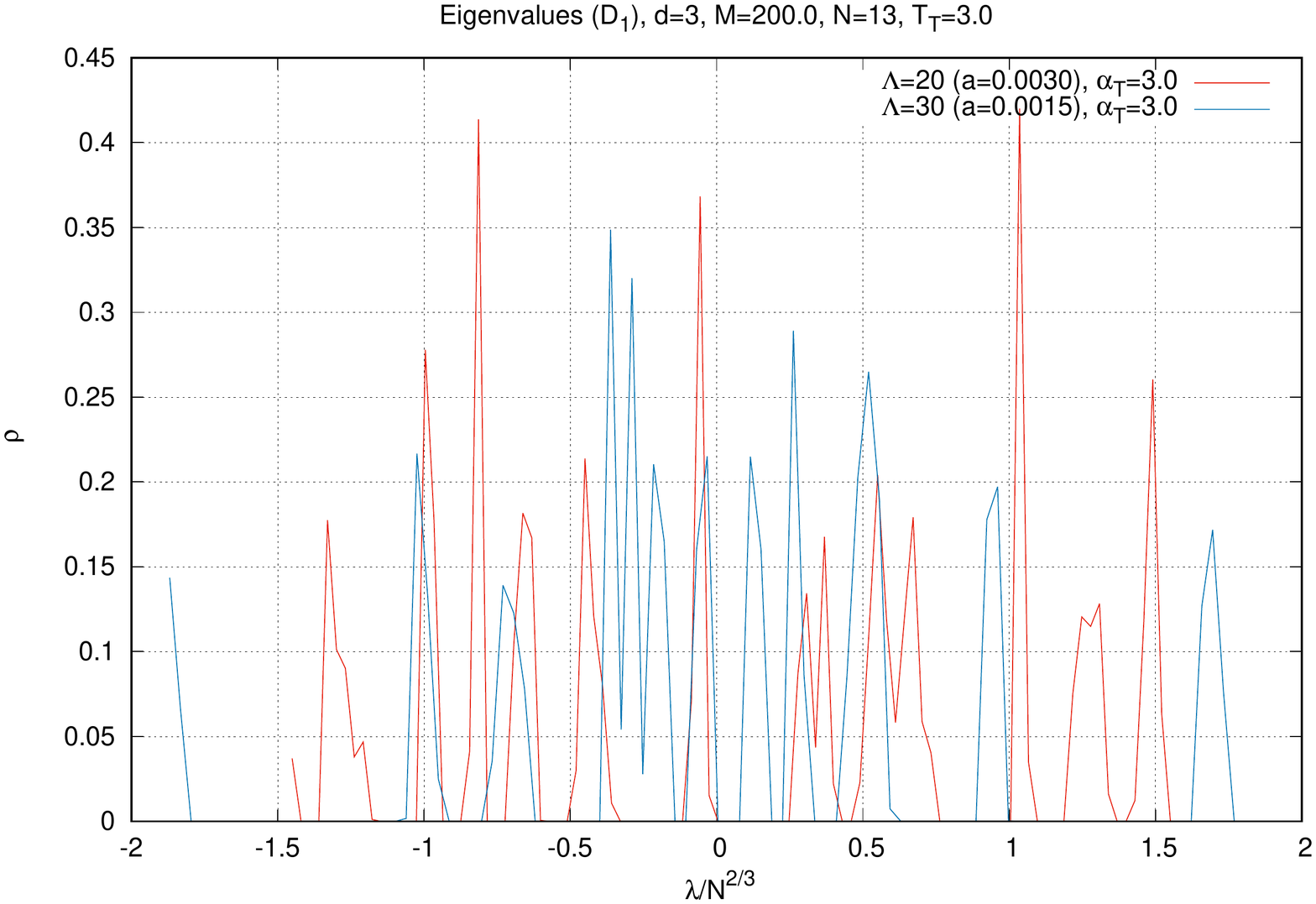}
  \includegraphics[width=10cm,angle=-0]{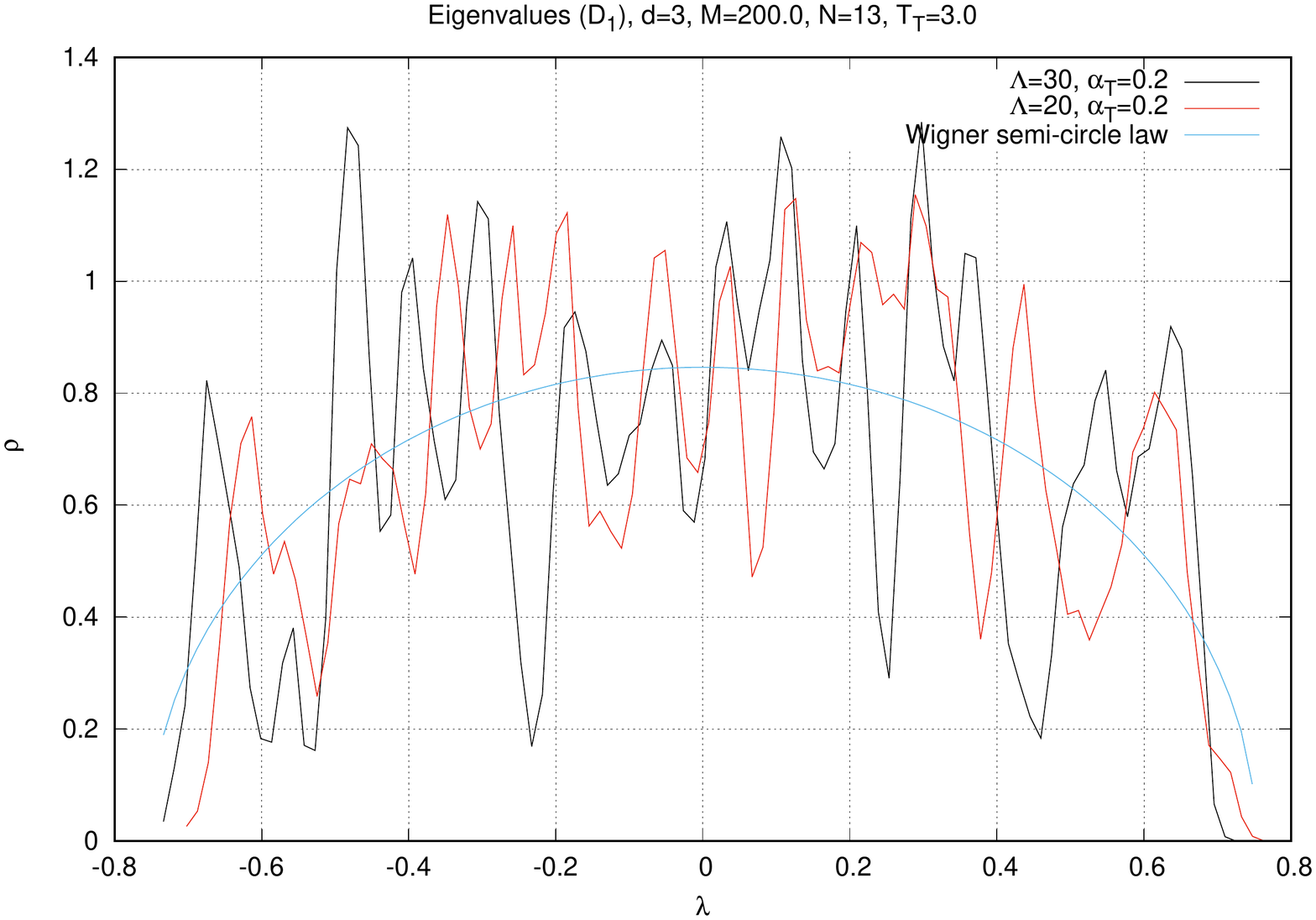}
\end{center}
\caption{The  eigenvalue distribution $\rho(\lambda)$ of the matrices $D_a$  for $\tilde{T}=3$ and $N=13$.}\label{sample3}
\end{figure}

\begin{figure}[htbp]
\begin{center}
  \includegraphics[width=10cm,angle=-0]{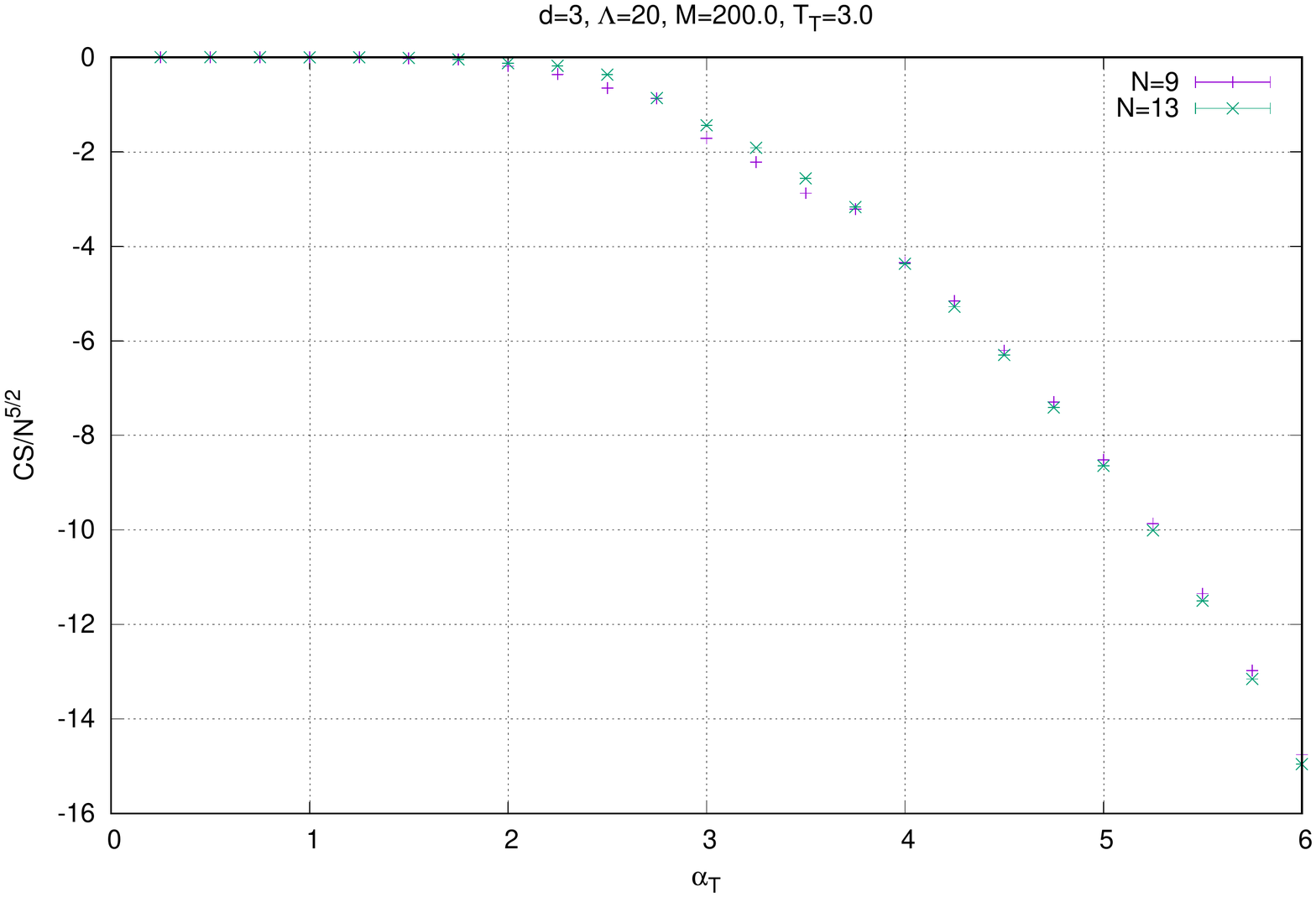}
  \includegraphics[width=10cm,angle=-0]{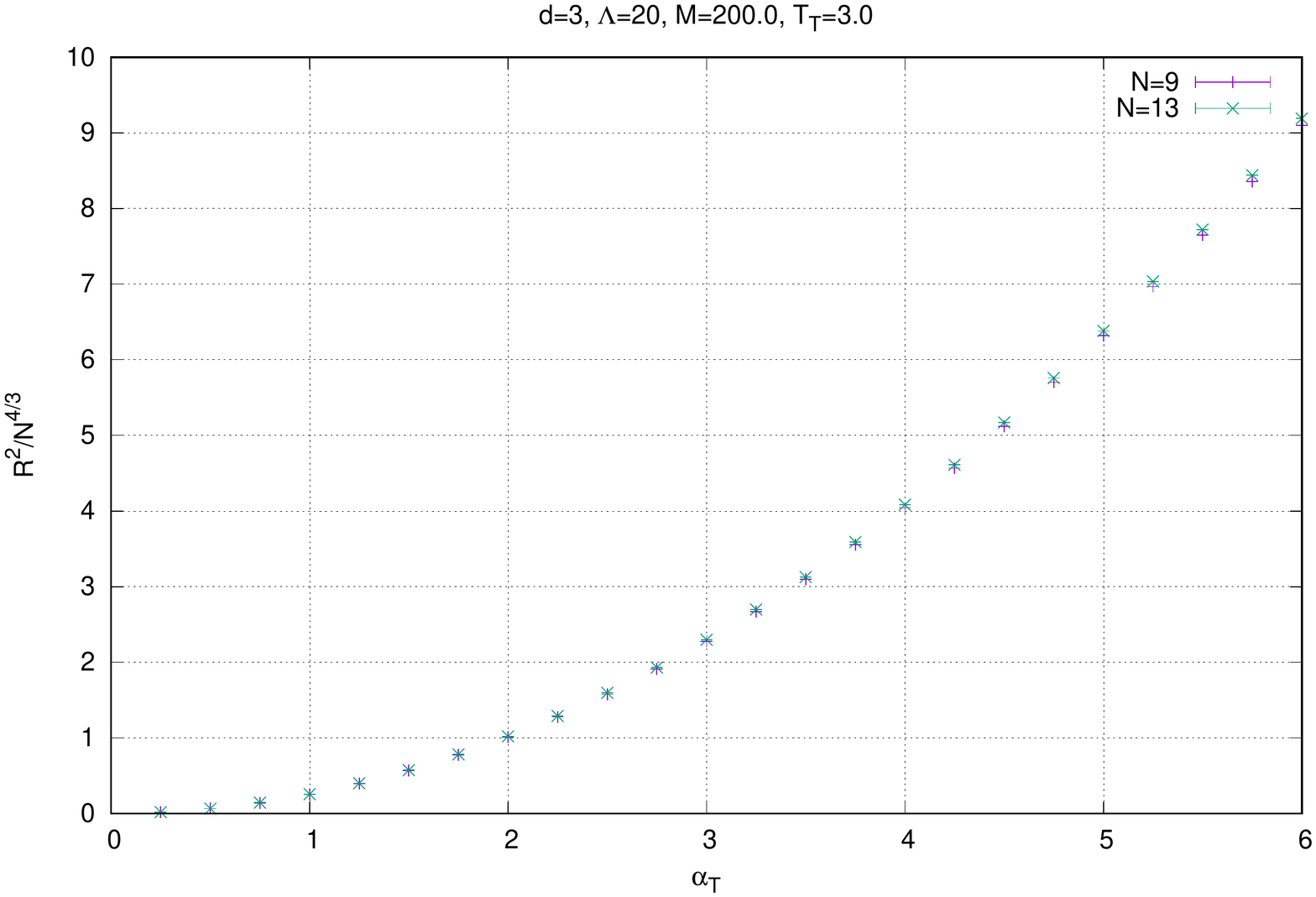}
  \includegraphics[width=10cm,angle=-0]{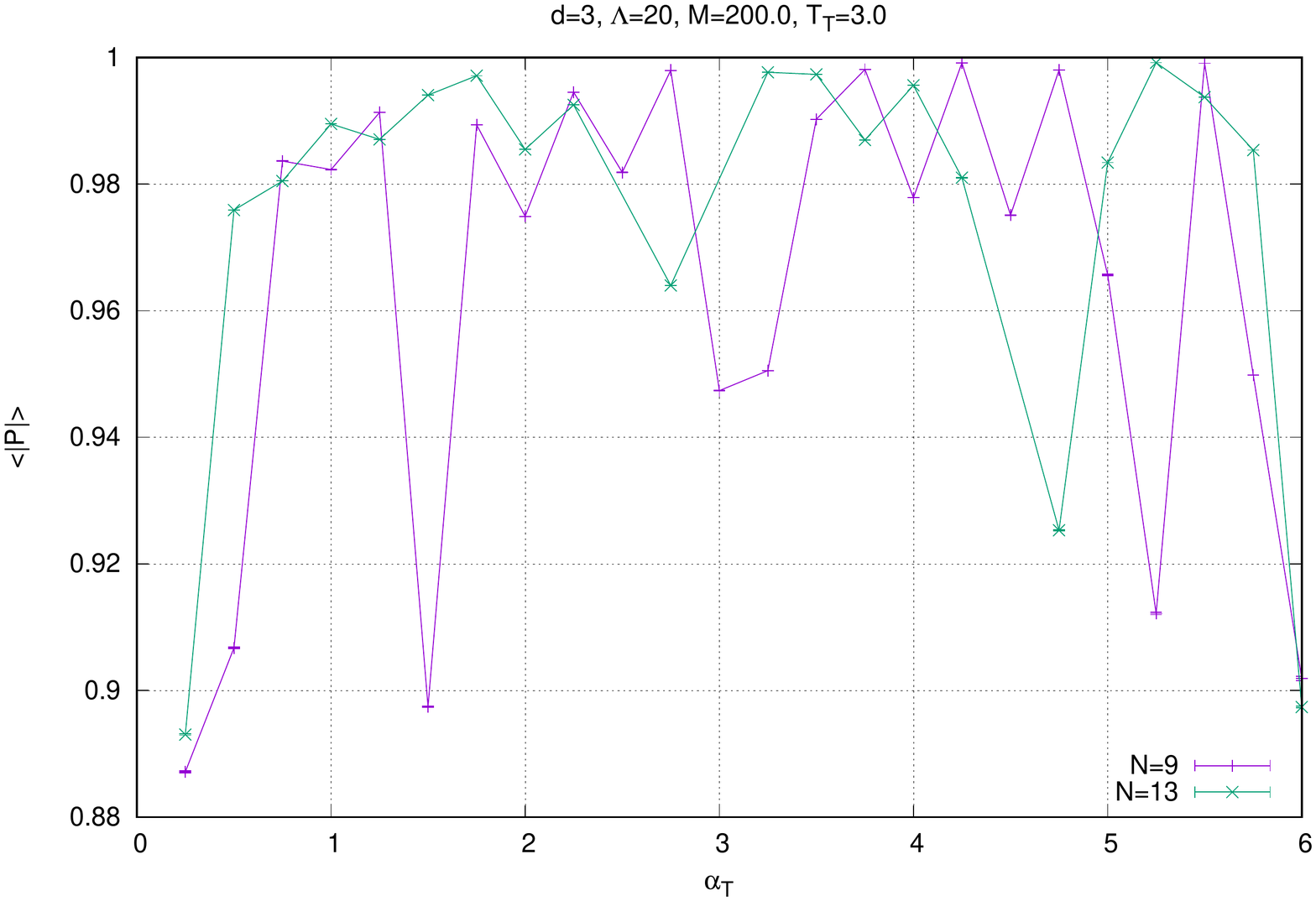}
\end{center}
\caption{The observables $\langle |P|\rangle$, $R^2$, ${\rm CS}$ for $\tilde{T}=3$ and $\Lambda=20$. }\label{sample4}
\end{figure}

As we have already remarked thermalization becomes more difficult with increasing matrix and lattice sizes $N$ and $\Lambda$ at high temperatures (and high $\tilde{\alpha}$).  The eigenvalue distribution $\rho(\lambda)$ and the observables  $\langle |P|\rangle$, $R^2$, ${\rm CS}$ for $\tilde{T}=3$ are shown in figure (\ref{sample3}) and (\ref{sample4}) \footnote{In the sample (\ref{sample3}) we run the Metropolis algorithm for $2^{q}+2^{13}$ steps where in most cases $q=15$ for $\Lambda=20$ and $q=16$ for $\Lambda=30$. Again the first $2^{q}$ steps  are discarded whereas histograms and averages are computed using the last $2^{13}$ steps. In the sample (\ref{sample4}) we run the Metropolis algorithm for $2^{15}+2^{13}$ steps and use only the lattice $\Lambda=20$.}. In particular, it is seen from the result of the Polyakov line that the baby fuzzy sphere exists always in the deconfined phase confirming the previous result.

We observe in the eigenvalue distributions shown on figure (\ref{sample3}) how we go slowly (as we decrease the value of $\tilde{\alpha}$) from the three-cut solution (the baby fuzzy sphere phase) to the multi-cut phase or uniform distribution (medium values of $\tilde{\alpha}$ in the Yang-Mills phase) to the Wigner semi-circle law (low $\tilde{\alpha}$ in the Yang-Mills phase). The transition point for $\tilde{T}=3$ and $N=13$ for both lattices $\Lambda=30$ and $\Lambda=20$ is found to be given by $\tilde{\alpha}=3.5\pm 0.5$ which is also consistent with  the one-loop estimation (\ref{oneloop}).

Let us also remark that in the Wigner-semi-circle law used to fit the data for $\tilde{\alpha}=0.5$ on  figure (\ref{sample3}) the fit includes the subleading $1/M$ correction which is very important at high temperatures (see equation (\ref{law3})).

Finally we note that for high temperatures $\tilde{T}$ and high gauge coupling constant $\tilde{\alpha}$ (deep inside the baby fuzzy sphere phase) it is observed that the three-cut configuration is lacking one of its two largest cuts indicating possibly the spontaneous breakdown of $SU(2)-$symmetry in the baby fuzzy sphere phase. However, this result is not conclusive from our current data and thus we will not contemplate its physical significance at this stage. 

In summary we obtain the phase diagram for $N=13$ and $\Lambda=20$, $\Lambda=30$ and $\Lambda=35$ ($a=0.05$) shown in figure (\ref{sample6}).
\subsection{The one-loop effective potential}
In the high temperature limit $T\longrightarrow \infty$ or equivalently $\beta\longrightarrow 0$ the Yang-Mills quantum mechanics $S_0+S_M$ given by (\ref{ac0})+(\ref{ac1}) together with (\ref{ac2}) reduces to a Yang-Mills matrix model given by
\begin{eqnarray}
S_1=NTr\bigg(-\frac{1}{2}[X_4,X_a]^2+\gamma^2X_a^2+\frac{2i\gamma}{3}\epsilon_{abc}X_aX_bX_c\bigg)+M(Tr X_a^2-\gamma^2c_2N)^2.\label{s1}
\end{eqnarray}
The fields $X_4$ and $X_a$ are given by $X_4=\beta^{1/4}A$ and $X_a=\beta^{1/4}\Phi_a$ while $\gamma$ is the only independent parameter in the model in the large $T$ limit given in terms of the temperature $T$ and the gauge coupling constant $\alpha$ by the combination
\begin{eqnarray}
\gamma=\beta^{1/4}\alpha.
\end{eqnarray}
Here we can show more explicitly that the mass term is suppressed in the high temperature limit as follows. We scale the fields as $X_a=\gamma D_a$ to obtain the action
\begin{eqnarray}
S_1=-\frac{N\gamma^2}{2}Tr[X_4,D_a]^2+N\gamma^4Tr\bigg(D_a^2+\frac{2i}{3}\epsilon_{abc}D_aD_bD_c\bigg)+M\gamma^4(Tr D_a^2-c_2N)^2.\label{s1v1}
\end{eqnarray}
Thus, if we keep the gauge coupling constant $\bar{\gamma}^4=N\gamma^4=\tilde{\alpha}^4/(\tilde{T}N)$ fixed in the large $N$ limit (which is here equivalent to the large $\tilde{T}$ limit) then $M\gamma^4=M\bar{\gamma}^4/N$ is suppressed in the large $N$ limit.

We know from previous work that the effective potential of similar Yang-Mills matrix models will involve a logarithmic correction \cite{CastroVillarreal:2004vhF}. We only need to determine its coefficient by performing a one-loop background field computation. Thus, we start by expanding the fields as
\begin{eqnarray}
X_a=B_a+Q_a~,~X_4=B_4+Q_4.
\end{eqnarray}
The background of interest is of the form
\begin{eqnarray}
B_4=0~,~  B_a=\mu \bigg(\begin{array}{cc}
       (J_a)_{n\times n} & 0 \\
      0 & b_a
      \end{array}\bigg).\label{fluct}
\end{eqnarray}
The generators $J_a$ transform in the spin $(n-1)/2$ irreducible representation of $SU(2)$ while the commuting matrices $b_a$ are assumed to transform in the reducible representation $0\oplus 0\oplus...\oplus 0$ ($N-n$ times).

We compute the quadratic action (the linear terms vanish by the equations of motions)
\begin{eqnarray}
  S_1[X]&=&S_1[B]+NTr\bigg[2[B_a,B_4][Q_4,Q_a]+[B_4,Q_4][B_a,Q_a]-\frac{1}{2}[B_4,Q_a]^2-\frac{1}{2}[B_a,Q_4]^2\nonumber\\
    &-&\gamma^2(2c_2M-1)Q_a^2+2i\gamma\epsilon_{abc}B_cQ_aQ_b+\frac{M}{N}[B_a,Q_a]^2+\frac{2M}{N}Q_aB_b^2Q_a+\frac{4M}{N}Q_aB_aB_bQ_b\bigg].\nonumber\\
\end{eqnarray}
This reads in the chosen background
\begin{eqnarray}
  S_1[X]&=&S_1[B]+NTr\bigg[-\frac{1}{2}[B_a,Q_4]^2-\gamma^2(2c_2M-1)Q_a^2+2i\gamma\epsilon_{abc}B_cQ_aQ_b\nonumber\\
    &+&\frac{M}{N}[B_a,Q_a]^2+\frac{2M}{N}Q_aB_b^2Q_a+\frac{4M}{N}Q_aB_aB_bQ_b\bigg].
\end{eqnarray}
We add the gauge-fixing and Faddeev-Popov ghost terms (after imposing a Lorentz gauge condition $[B_{\mu},Q_{\mu}]=0$)
\begin{eqnarray}
  \Delta S_1[X]&=&-\frac{1}{\xi}Tr [B_a,Q_a]^2+Tr c^{\dagger}[B_a,[B_a,c]].
\end{eqnarray}
We choose the Landau-like gauge $\xi=1/M$ which goes to $0$ when $M\longrightarrow\infty$. After integrating out the fluctuations $Q_a$ and $Q_4$ we get the effective action ( effective potential)
\begin{eqnarray}
  V_{\rm eff}&=&S_1[B]+\frac{1}{2}TR\log {\cal B}_a^2+\frac{1}{2}tr_3TR\log \Omega-TR\log{\cal B}_a^2.
\end{eqnarray}
The second term comes from the integration of $Q_4$, the third term comes from the integration of $Q_a$ whereas the fourth term comes from the integration of the ghost fields. The operator $\Omega$ is given by 
\begin{eqnarray}
 \Omega=\gamma^2\delta_{ab}-(M-\frac{1}{\xi}){\cal B}_a{\cal B}_b-2i\gamma\epsilon_{abc}B_c+\frac{2M}{N}(B_c^2-\gamma^2c_2N)\delta_{ab}+\frac{4M}{N}B_aB_b.
\end{eqnarray}
We get then the effective potential 
\begin{eqnarray}
  V_{\rm eff}&=&N\hat{c}_2n(\gamma^2\mu^2-\frac{2\gamma}{3}\mu^3)+M(\mu^2\hat{c}_2n-\gamma^2c_2N)^2-{\cal N}\log \mu+\Delta V_{\rm eff}.
\end{eqnarray}
Three assumptions are made here:
\begin{itemize}
\item The quantum correction $\Delta V_{\rm eff}$ which depends on the Laplacian $\Omega$  is negligible compared to the logarithmic potential which provides therefore the main quantum correction to the classical potential (the first two terms).
\item Furthermore, the number of degrees of freedom ${\cal N}$ is simply equal $N^2$. Here we are assuming that the commuting matrices $b_a$ in (\ref{fluct}) have eigenvalues around $0$ which are negligible compared to the eigenvalues of $J_a$.
\item Also, we are assuming here the Landau-like gauge $\xi=1/M$. But generic gauges seems to have only a mild effect on the result.
  \end{itemize}
These three assumptions  will be confirmed by Monte Carlo simulation shortly.

The classical equation of motion is
\begin{eqnarray}
  \gamma-\mu+\frac{2M}{N\gamma}(\mu^2\hat{c}_2n-\gamma^2c_2N)=\frac{{\cal N}}{2\hat{c}_2nN\gamma}\frac{1}{\mu^2}.
\end{eqnarray}
The solution as a $1/M$ expansion reads
\begin{eqnarray}
  \mu=\mu_0+\frac{\mu_1}{M}+...\label{qs1}
\end{eqnarray}
\begin{eqnarray}
  \mu_0^2=\frac{\gamma^2c_2N}{\hat{c}_2n}~,~\mu_1=\frac{{\cal N}}{8\hat{c}_2^2n^2}\frac{1}{\mu_0^3}+\frac{N\gamma}{4\hat{c}_2n}(1-\frac{\gamma}{\mu_0}).\label{qs2}
\end{eqnarray}
Higher order corrections in $1/M$ are irrelevant here. Furthermore, with the rescalings $\mu=\gamma a$, $\tilde{\gamma}^4=Nn\hat{c}_2\gamma^4/{\cal N}$ and $\tilde{M}=M/(Nn\hat{c}_2)$ we rewrite this potential as
\begin{eqnarray}
  \frac{V_{\rm eff}}{{\cal N}}&=&\tilde{\gamma}^4(a^2-\frac{2}{3}a^3)+\tilde{M}\tilde{\gamma}^4(a^2\hat{c}_2n-c_2N)^2-\log a.\label{Mterm}
\end{eqnarray}
The effective scale $a$ is given by the solution of the equation of motion $V^{\prime}_{\rm eff}=0$. The condition  $V^{\prime\prime}_{\rm eff}=0$ tell us when we go from a bounded potential (the solution $a$ is a minimum) to an unbounded potential ($a$ becomes a maximum). The critical value of the scale $a$  is found to be given by 
\begin{eqnarray}
  a_*=\frac{3}{16\tilde{M}n^2\hat{c}_2^2}\bigg[1-\sqrt{1+\frac{64\tilde{M}n^2\hat{c}_2^2}{9}(2\tilde{M}c_2\hat{c}_2Nn-1)}\bigg].
\end{eqnarray}
If we set $\tilde{M}=0$ (the model without a regularization or by using the high temperature scaling (\ref{ol})) we get the critical value $a_*=2/3$.  By replacing in  $V^{\prime}_{\rm eff}=0$ or  $V^{\prime\prime}_{\rm eff}=0$ we get in this case the following critical values

\begin{eqnarray}
\frac{8}{27}\tilde{\gamma}^4=1\Rightarrow \gamma^4|_*=\frac{9}{4}N\Rightarrow   \tilde{\alpha}^4_*=\frac{9}{4}\tilde{T}N^3.\label{oneloop0}
\end{eqnarray}
This suggest that the baby fuzzy sphere phase is pushed with increasing $N$ to higher values of the gauge coupling constant $\tilde{\alpha}$, i.e. it is not stable. But recall that the model without a regularization $\tilde{M}=0$  is unstable so this result should be taken with a grain of salt. On the other hand, this result also suggests that  the correct scaling in the Gaussian approximation of matrix Yang-Mills quantum mechanics is slightly different from (\ref{419})  given in fact by 
\begin{eqnarray}
\tilde{\alpha}=\alpha N^{1/3}~,~\bar{T}=\tilde{T}N^3.
\end{eqnarray}
Now, there is a problem with the $M-$term in (\ref{Mterm}) as it stands. It is not difficult to see that this term (coming from the double-trace term $(Tr (X_a^2-\gamma^2c_2))^2$ which kills off only the zero mode of the normal scalar field in the large $M$ limit as suggested by equation (\ref{ol})) can also be obtained from the much stronger potential $Tr(X_a^2-\gamma c_2)^2$ which kills the whole normal scalar field. As we have seen from equation (\ref{Bmass}) the (intended) effect of the double-trace term is really only to modify the mass term by effectively adding to it a correction which depends  on the second moment or radius $Tr\Phi_a^2/N=R^2$. A better treatment of the M-term is then to use the quantum solution  given by equations (\ref{qs1}) and (\ref{qs2}) to get the analogue of the approximation (\ref{beauty}) in the baby fuzzy sphere phase. The result can be rewritten as (we simply set $X_a=B_a$ in this approximation)
\begin{eqnarray}
TrX_a^2-Nc_2\gamma^2=\frac{N\Delta}{M}+O(\frac{1}{M^2}).\label{beauty1}
\end{eqnarray}
The $\Delta$ is given by 
\begin{eqnarray}
  \Delta &=&\frac{1}{2}\gamma(\mu_0-\gamma)+\frac{{\cal N}}{4\hat{c}_2nN}\frac{1}{\mu_0^2}\nonumber\\
  &=&\frac{\alpha^2}{2\sqrt{T}}\big(\pm \frac{N^{3/2}}{2\sqrt{\hat{c}_2n}}-1\big)+\frac{\sqrt{T}}{\alpha^2}\frac{{\cal N}}{N^4}.
\end{eqnarray}
This should be compared with the analogous result in the Yang-Mills phase (obtained by means of the approximation (\ref{beauty})) which reads

\begin{eqnarray}
TrX_a^2-Nc_2\gamma^2=\frac{N\Delta}{M}+O(\frac{1}{M^2})~,~\Delta=\frac{\alpha^2}{2\sqrt{T}}\big(\frac{T}{\alpha^4}\frac{2}{N^4}-1\big)+\frac{\sqrt{T}}{\alpha^2}\frac{1}{N^2}.
\end{eqnarray}
By the assumption and observation (in Monte Carlo simulation) of the continuity of the radius $R^2= Tr\Phi_a^2/N$ we can see that the number of degrees of freedom ${\cal N}$ is indeed ${\cal N}=N^2$. The quantum term ${\sqrt{\tilde{T}}}/\tilde{\alpha^2}$ is only relevant at small values of $\tilde{\alpha}$ where the Yang-Mills phase exists whereas at large values of $\tilde{\alpha}$ which dominate the baby fuzzy sphere phase at large temperatures this term is negligible. At the critical boundary between the baby fuzzy sphere phase and the Wigner's semi-circle law we should have (we observe that the positive sign is the preferred solution)
\begin{eqnarray}
\tilde{\alpha}_*^4=\frac{(2\sqrt{6})^{1/4}}{N^{3/8}}\tilde{T}^{1/4}.\label{cont}
\end{eqnarray}
This is the boundary between the Yang-Mills phase at small values of $\tilde{\alpha }$ where the Wigner's semi-circle law dominates and the baby fuzzy sphere phase. At large values of $\tilde{\alpha }$ the transition from  the baby fuzzy sphere phase to the Wigner's semi-circle law is not immediate but goes through the phase of commuting matrices with a uniform distribution and the estimation of the critical boundary requires the use of the effective potential (\ref{Mterm}) as we will do next.

But first the result (\ref{beauty1}) should also be compared with Monte Carlo simulation of the radius $R^2$ given more exactly by (\ref{obs2}). More precisely, we measure

\begin{eqnarray}
  \tilde{\Delta}=\frac{M(R^2-c_2\alpha^2)}{\alpha^2}.
\end{eqnarray}
The one-loop result from (\ref{beauty1}) is clearly given by $\tilde{\Delta}=\Delta/\gamma^2=\sqrt{T}\Delta/\alpha^2$ which is equal to $a+b/x^4$ where $x\equiv \tilde{\alpha}^4$, $b\equiv\tilde{T}$ and $a$ sales as $a/N^{3/2}=\pm 1/(2\sqrt{6})$ in the large $N$ limit.

\begin{figure}[htbp]
\begin{center}
  \includegraphics[width=10cm,angle=-0]{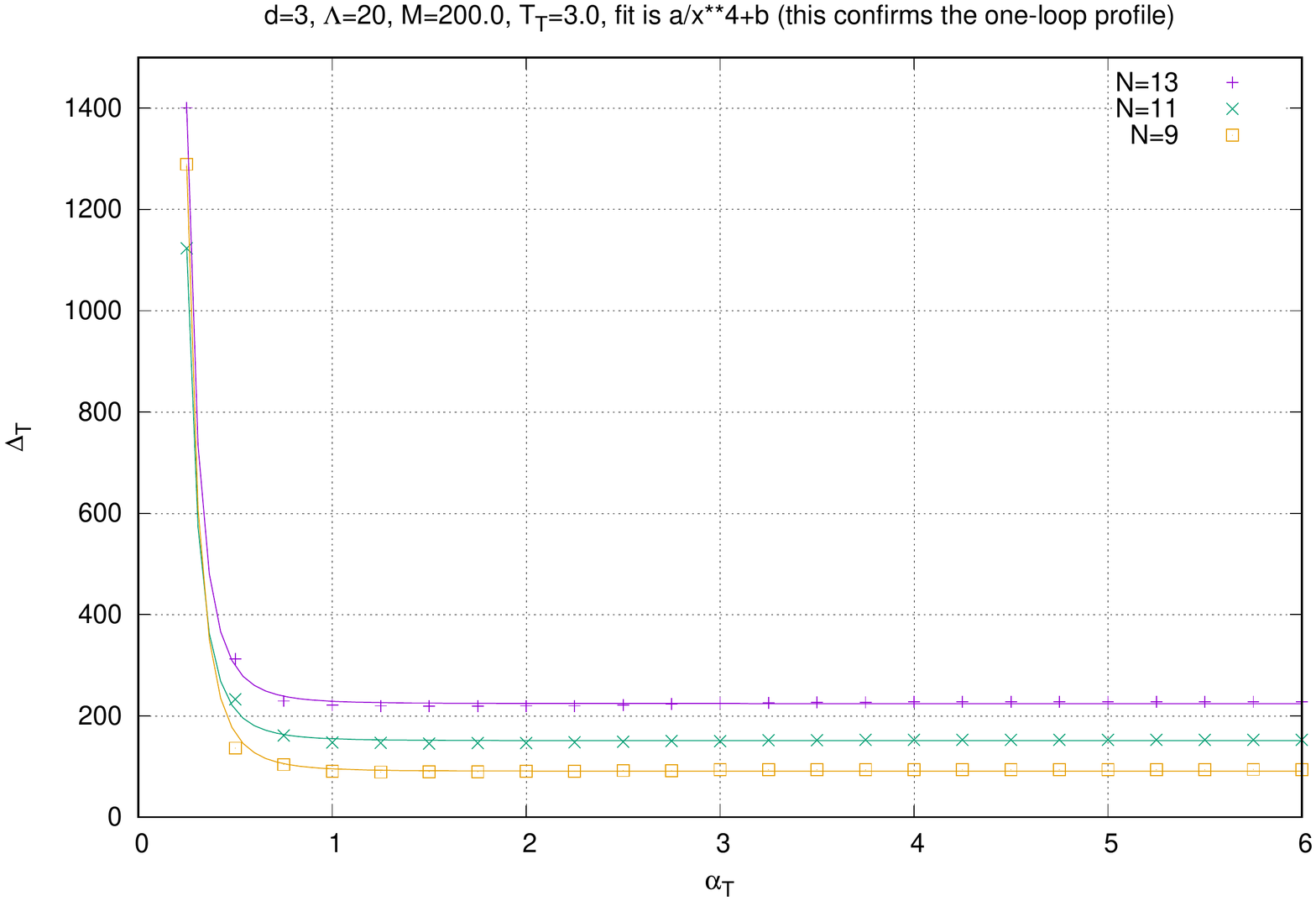}
  \includegraphics[width=10cm,angle=-0]{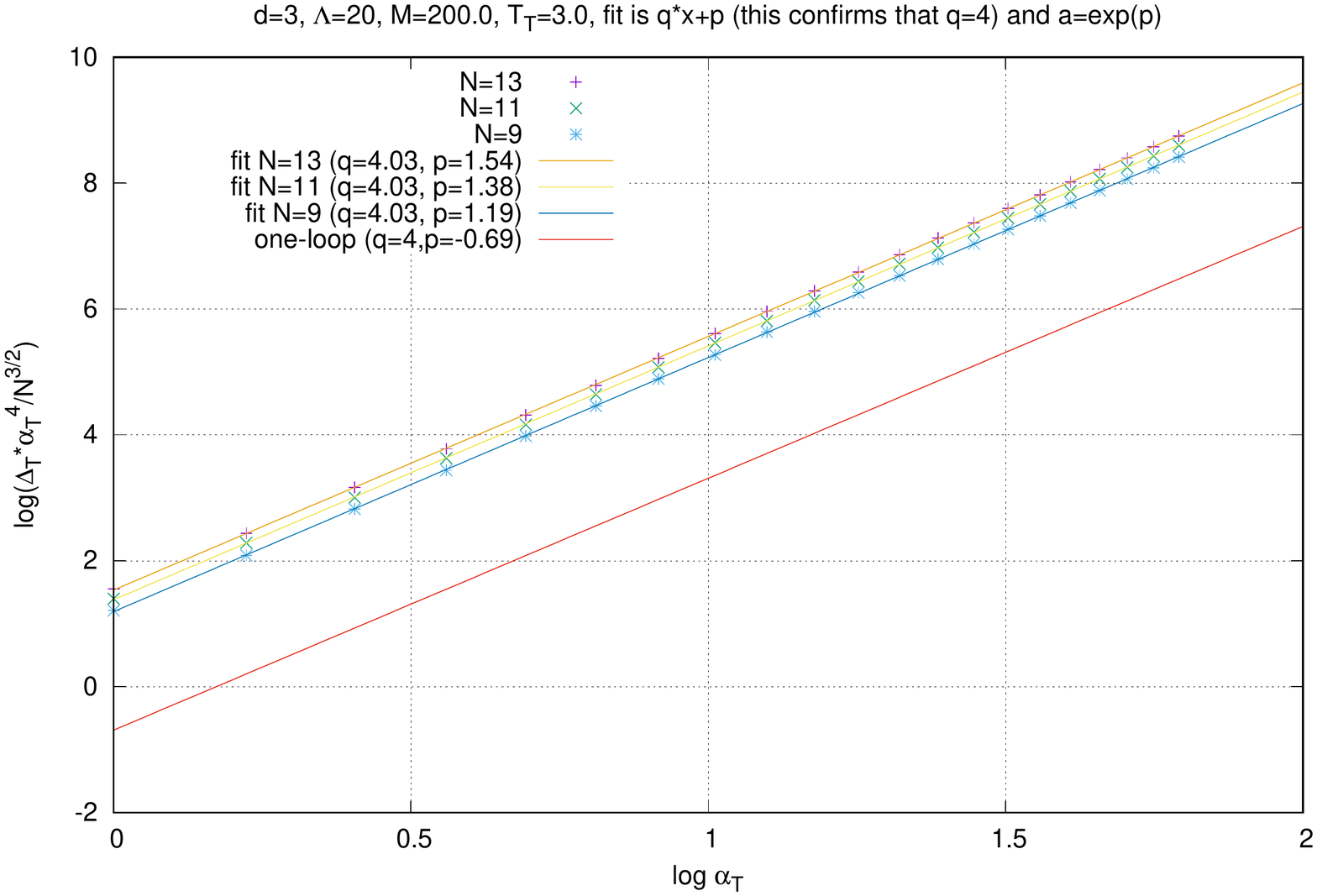}
  \includegraphics[width=10cm,angle=-0]{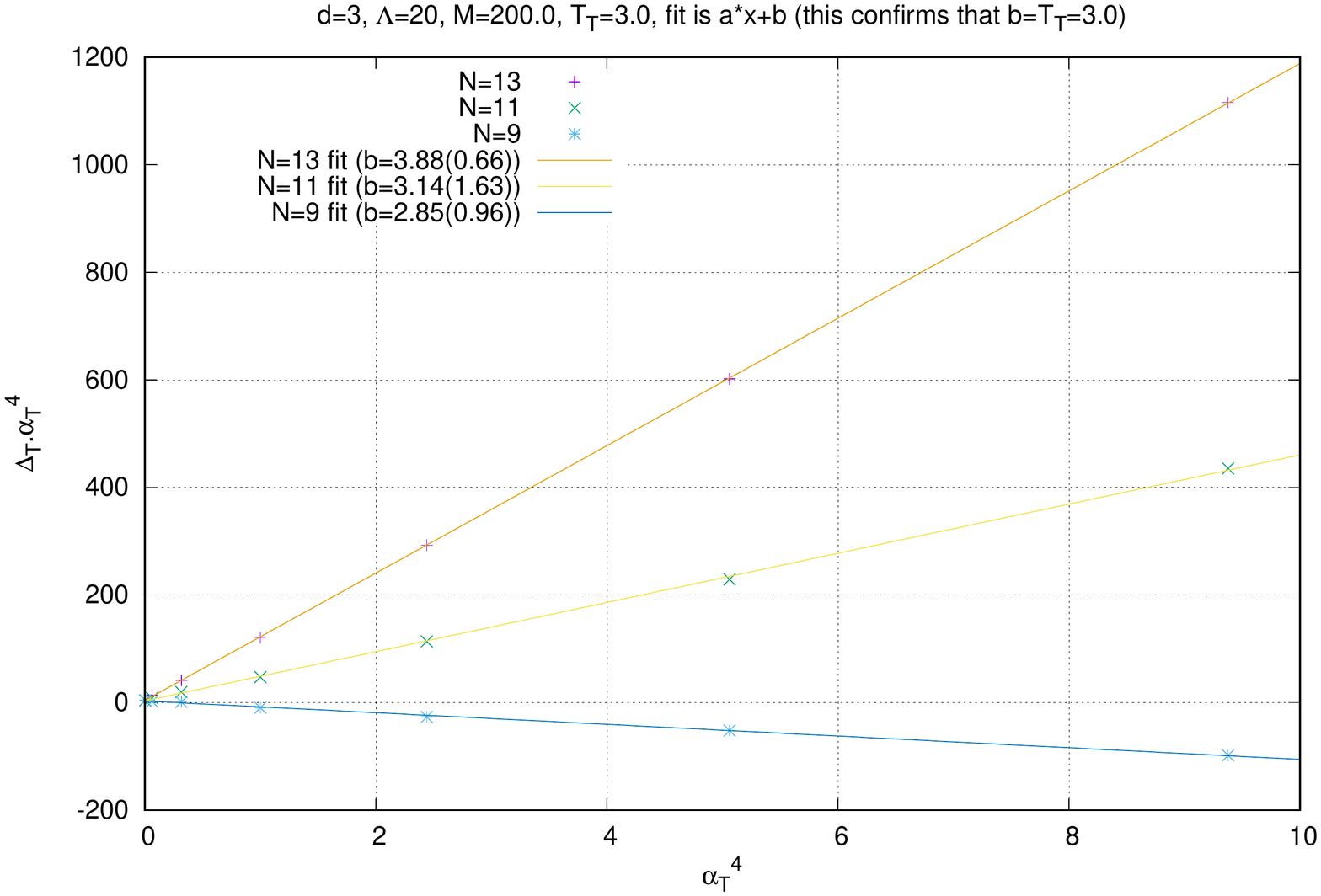}
\end{center}
\caption{Monte Carlo tests of the one-loop result (\ref{beauty1}). }\label{sample7}
\end{figure}

The Monte Carlo results for the observable $\tilde{\Delta}$ are shown on figure (\ref{sample7}). The quantum correction $b/\tilde{\alpha^4}$ is also confirmed by the Monte Carlo result (and thus it is confirmed by three different methods). Thus, for large values of the gauge coupling constant  $\tilde{\alpha}$ (where the baby fuzzy sphere phase occurs at high temperatures) this quantum term vanishes as indicated by the first graph on figure (\ref{sample7}).  For small values of the gauge coupling constant where the Yang-Mills phase exists this quantum term dominates with precisely the value $b=\tilde{T}$. Indeed,  the slope $q$ in the second graph while the intercept $b$ in the third graph are found to be consistent with the values $q=4$ and $b=\tilde{T}$.

The classical piece $a/N^{3/2}=\pm 1/(2\sqrt{6})$ shows however a discrepancy with the exact result given by Monte Carlo simulations as indicated by the intercept $p=\log a$ in the second graph. This discrepancy will be reflect as a discrepancy between the critical boundary as derived from the effective potential (\ref{Mterm}) and the critical boundary measured in Monte Carlo simulations by the eigenvalue distribution $\rho(\lambda)$.  

By using now the approximation (\ref{beauty1}) (in a way similar to mean field approximation) the relevant action becomes
\begin{eqnarray}
S_1=NTr\bigg(-\frac{1}{2}[X_4,X_a]^2+\gamma^2\epsilon X_a^2+\frac{2i\gamma}{3}\epsilon_{abc}X_aX_bX_c\bigg).
\end{eqnarray}
Thus, we are effectively treating the $M-$term as suppressed as it should be in the high temperature limit without (hopefully) destroying the correct behavior of the solution near the critical boundary (this also shows that large $N$ and large $T$ are somehow equivalent in this scheme as suggested by (\ref{s1v1})). The parameter $\epsilon$ is given by 
\begin{eqnarray}
\epsilon \equiv 1+\frac{\Delta}{\gamma^2}=\frac{1}{2}\pm \frac{N^{3/2}}{4\sqrt{\hat{c}_2n}}+\frac{1}{\gamma^4}\frac{{\cal N}}{N^4}.\label{qsd}
\end{eqnarray}
This looks as if $\epsilon$ is dominated by the second (classical) term but near the critical boundary the third (quantum) term is in fact of a comparable size. After some more calculation we get now the effective potential

\begin{eqnarray}
\boxed{ \frac{V_{\rm eff}}{{\cal N}}=\tilde{\gamma}^4(\epsilon a^2-\frac{2}{3}a^3)-\log a.}
\end{eqnarray}
We get the critical points
\begin{eqnarray}
\boxed{a_*=\frac{2\epsilon}{3}.}
\end{eqnarray}
\begin{eqnarray}
  \frac{8}{27}\tilde{\gamma}^4\epsilon^3=1\Rightarrow \gamma^4\epsilon^3|_*=\frac{9}{4}N.
\end{eqnarray}
In terms of $\tilde{\alpha}_*$ the critical boundary reads 
\begin{eqnarray}
  \boxed{\tilde{\alpha}_*=\bigg(\frac{2\sqrt{6}}{2\sqrt{6} x\mp 1}\bigg)^{1/4}\frac{\tilde{T}^{1/4}}{N^{3/8}}.}\label{oneloop}
\end{eqnarray}
The parameter $x$ is the solution of the depressed cubic equation $x^3-9x/4\pm 9/8\sqrt{6}=0$. The two signs $+$ and $-$ give solutions with opposite  signs   and  among these six possible solutions three are physical giving the positive values

\begin{eqnarray}
  \bigg(\frac{8\sqrt{6}}{2\sqrt{6} x\mp 1}\bigg)^{1/4}=3.90, 0.95, 0.86.\label{390}
\end{eqnarray}
The largest value  is the one consistent with the quantum correction given by the third term in (\ref{qsd}) being negligible.

The one-loop result (\ref{oneloop})  is larger than the continuity result (\ref{cont}) as it should be since the phase of commuting matrices with a uniform distribution occurs at higher values of the gauge coupling constant $\tilde{\alpha}$ than the Wigner's semi-circle law. Both results suggest that the baby fuzzy sphere phase is pushed with increasing $N$ to lower values of the gauge coupling constant $\tilde{\alpha}$, i.e. this phase is more stable. But with the fact that $\tilde{T}$ is large this effect can not be easily discerned in Monte Carlo simulations. 

\begin{figure}[htbp]
\begin{center}
  \includegraphics[width=20cm,angle=-0]{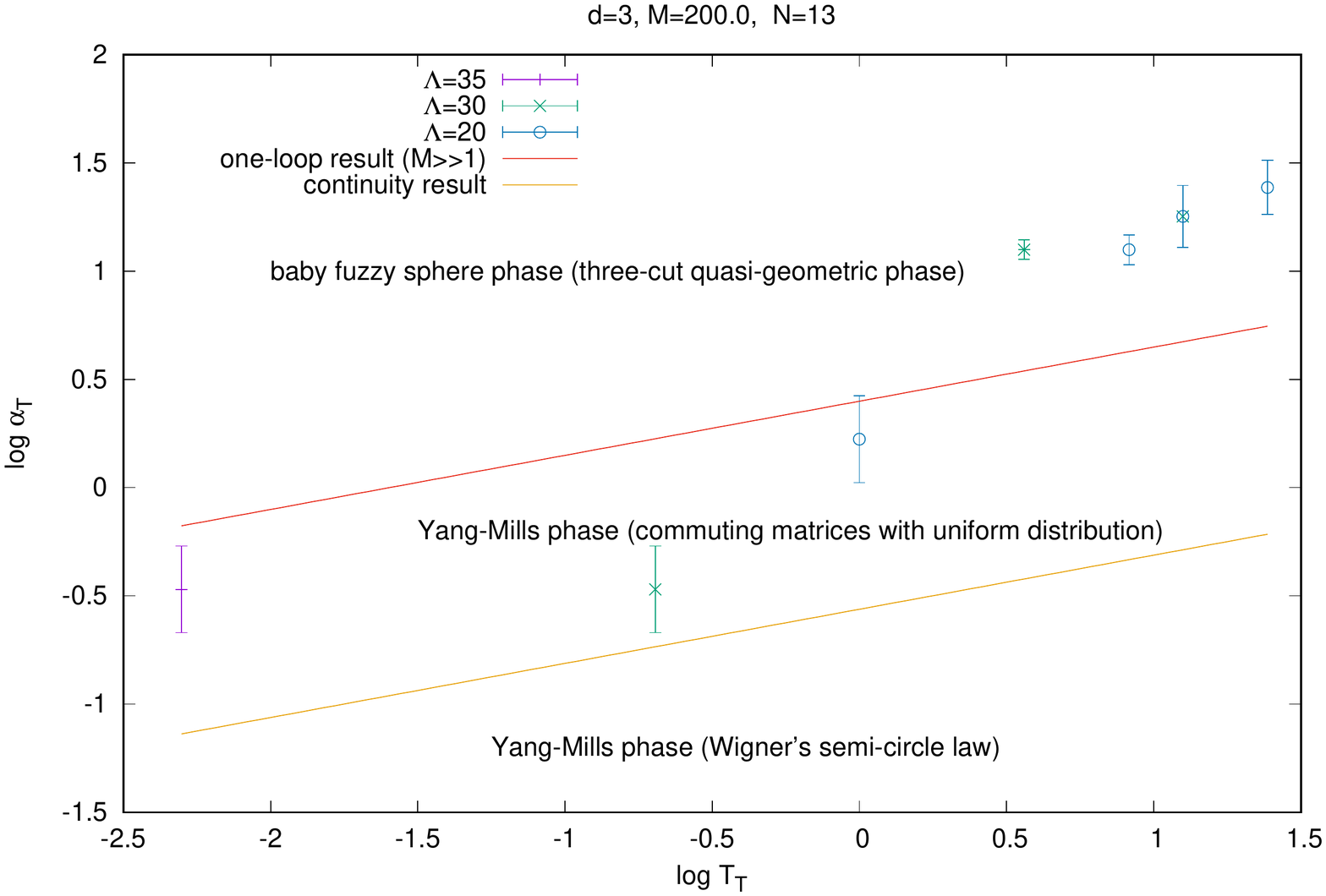}
\end{center}
\caption{The phase diagram for $N=13$. }\label{sample6}
\end{figure}

The Monte Carlo data are found to lie between the two lines (\ref{oneloop0}) and (\ref{oneloop}) but much closer to (\ref{oneloop}). There remains however a discrepancy between the Monte Carlo data and (\ref{oneloop}) as observed on the phase diagram (\ref{sample6}) which is due to the discrepancy between the one-loop  and exact results. This discrepancy is nicely illustrated in the observable $\tilde{\Delta}$ which encodes the next-to-leading-order correction in $1/M$ of the radius $R^2$.

\subsection{Hagedorn transition in the Yang-Mills phase}
\begin{figure}[htbp]
\begin{center}
  \includegraphics[width=15cm,angle=-0]{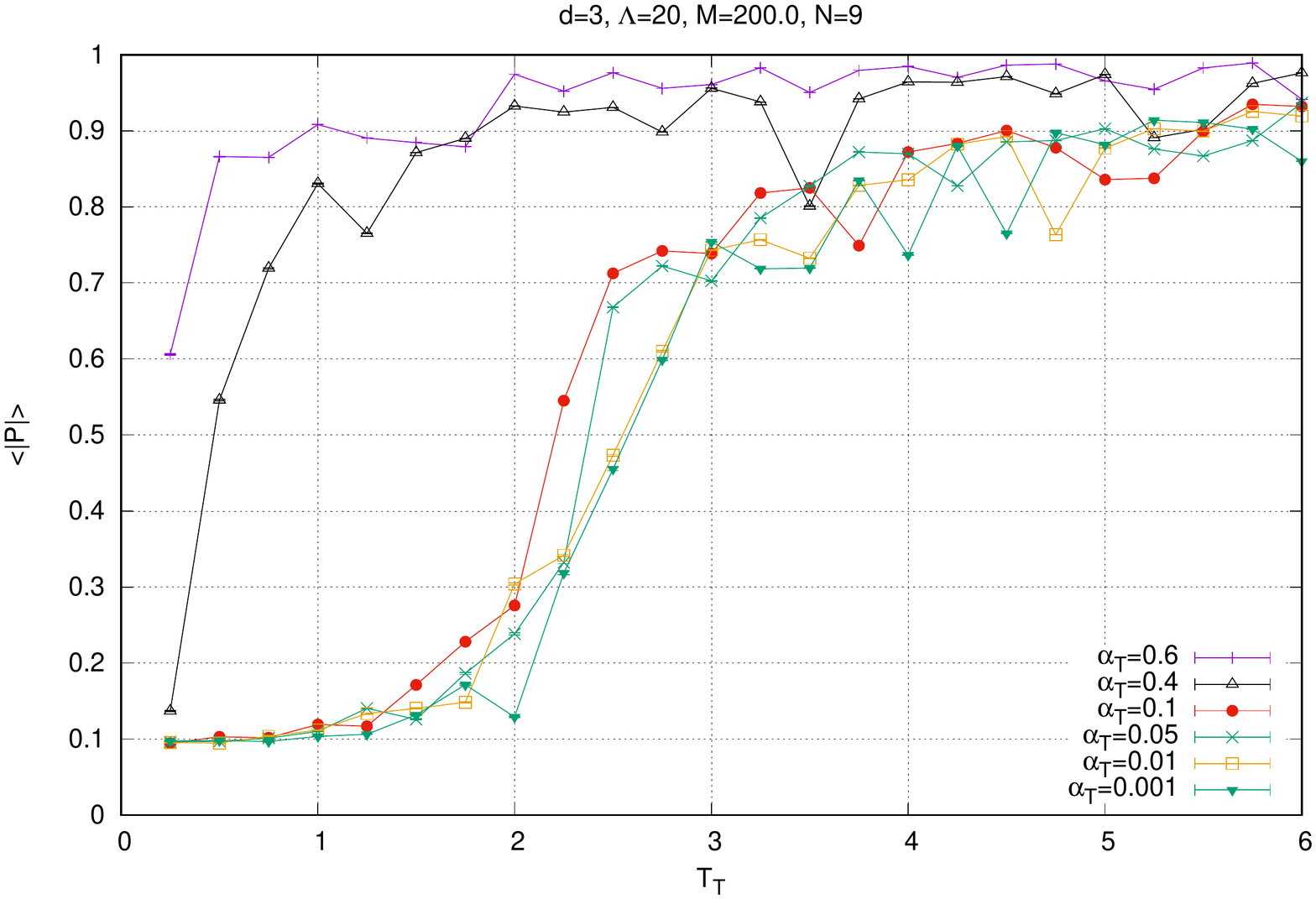}
\end{center}
\caption{The Hagedorn transition in the Yang-Mills phase. }\label{sample8}
\end{figure}

The Yang-Mills phase (where the Chern-Simons term is virtually zero) is also characterized by a Hagedorn transition which divides it into a confinement phase at low temperatures and a deconfinement phase at high temperatures (despite the fact that in the regularized action the squared mass $m^2=d^{2/3}$ has changed to $2\alpha^2$). However, the  Hagedorn temperature seems to depend on the radius of the Wigner's semi-circle law. At very low values of the gauge coupling constant $\tilde{\alpha}$ (below $\tilde{\alpha}=0.1$) where the radius of the Wigner's semi-circle law is given by (\ref{law2}) the  Hagedorn temperature occurs at the same value for all values of  $\tilde{\alpha}$. But for larger values of the gauge coupling constant $\tilde{\alpha}$ where the radius of the Wigner's semi-circle law is given by (\ref{law1}) the  Hagedorn temperature decreases with increasing  $\tilde{\alpha}$ until the confinement phase disappears altogether at the critical boundary with the baby fuzzy sphere phase. Indeed, the baby fuzzy sphere phase exists always in the deconfinement phase. See figure (\ref{sample8}) for a sample.

\section{Conclusion and outlook}
In this note we have attempted to unify two approaches of quantum gravity (the gauge/gravity duality and the matrix/geometry approach) in a single Gaussian/cubic matrix model given by the large $d$ approximation of the bosonic part of the M-(atrix) theory action supplemented with a mass deformation (given by the Chern-Simons term). The action is given by equation (\ref{ter2}) or equation (\ref{ac0}). The M-(atrix) theory in $d=9$ is precisely the BFSS matrix quantum mechanics.

The characteristics of the  black-hole-to-black-string phase transition are captured to a very good accuracy using the Gaussian model whereas only a remnant  of the Yang-Mills-to-fuzzy-sphere phase  is reproduced by means of the cubic action (regularized effectively by a double-trace potential which removes only the zero mode of the scalar sector).

For simplicity we have worked with $d=3$ in the second case whereas in the first case we have worked with $d=9$.

We have convinced ourselves by a combination of the Monte Carlo, perturbative and matrix methods that only a remnant of the fuzzy sphere solution (termed here the baby fuzzy sphere) persists in this approximation. This solution is a three-cut configuration corresponding to the irreducible representation of $SU(2)$ given by the direct sum $0 \oplus \frac{1}{2}$.

The Yang-Mills phase becomes on the other hand divided into two distinct regions with what seems to be a crossover between them. The first  region is dominated by the Wigner's semi-circle law at very low values of the gauge coupling constant $\tilde{\alpha}$ and a second region inside the Yang-Mills phase dominated by a uniform distribution occurring at medium values of the gauge coupling constant before reaching the baby fuzzy sphere boundary. 

The boundary between the two phases  is constructed and it is shown that the scaling of the gauge coupling constant and the temperature  in the two phases is possibly different resulting in the fact that either the geometric baby fuzzy sphere phase removes the uniform confining phase (which is the most plausible physical possibility) or the other way around, i.e. the gauge theory uniform confining phase removes the baby fuzzy sphere phase.   

This discrepancy (the fact that only a remnant "baby fuzzy sphere" configuration is found to be stable) is certainly not due to the double-trace regulator (\ref{ac1}) but it is due to the approximation (\ref{ter2}) of the large $d$ behavior of the bosonic part of the M-(atrix) theory action with a mass deformation which is given by equation (\ref{ter}). Indeed, the approximation (\ref{ter2}) of the action (\ref{ter}) is certainly valid for small values of $\alpha$ near $\alpha=0$ but the extrapolation to much larger values of $\alpha$ does not necessarily need to be valid but in fact it is expected to break down at some point.

However, studying the action (\ref{ter2}) with the double-trace potential (\ref{ac1}) in its own right warrants our interest and thus more Monte Carlo data (with larger lattices, larger matrix sizes and higher statistics) is currently being pursued.

\end{document}